\documentclass[fleqn,usenatbib]{mnras}

\usepackage{newtxtext,newtxmath}

\usepackage[T1]{fontenc}

\DeclareRobustCommand{\VAN}[3]{#2}
\let\VANthebibliography\thebibliography
\def\thebibliography{\DeclareRobustCommand{\VAN}[3]{##3}\VANthebibliography}

\usepackage{graphicx}	
\usepackage{amsmath}	
\usepackage{csvsimple}  
\usepackage{dirtytalk}  
\hypersetup{
	colorlinks=true,       
	linkcolor=blue,        
	citecolor=blue,        
	filecolor=magenta,     
	urlcolor=blue         
}
\usepackage{booktabs}   

\usepackage{siunitx}
\DeclareSIUnit \h {\ensuremath{\mathit{h}}}
\DeclareSIUnit \pc {pc}

\usepackage{xargs} 
\usepackage[pdftex,dvipsnames]{xcolor}
\usepackage[colorinlistoftodos,prependcaption,textsize=tiny]{todonotes}
\newcommandx{\unsure}[2][1=]{\todo[linecolor=red,backgroundcolor=red!25,bordercolor=red,#1]{#2}}
\newcommandx{\change}[2][1=]{\todo[linecolor=blue,backgroundcolor=blue!25,bordercolor=blue,#1]{#2}}
\newcommandx{\info}[2][1=]{\todo[linecolor=OliveGreen,backgroundcolor=OliveGreen!25,bordercolor=OliveGreen,#1]{#2}}
\newcommandx{\improvement}[2][1=]{\todo[linecolor=Plum,backgroundcolor=Plum!25,bordercolor=Plum,#1]{#2}}
\newcommandx{\addcite}[2][1=]{\todo[linecolor=yellow,backgroundcolor=yellow!25,bordercolor=yellow,#1]{Add cite}}

\title[Weak lensing magnification of Type Ia Supernovae]{Weak lensing magnification of Type Ia Supernovae from the Pantheon sample}

\author[P.Shah et al.]{
Paul Shah,$^{1}$\thanks{E-mail: paul.shah.19@ucl.ac.uk}
Pablo Lemos,$^{1, 2}$
Ofer Lahav$^{1}$
\\
$^{1}$Department of Physics and Astronomy, University College London, Gower Street, London, WC1E 6BT, UK\\
$^{2}$Department of Physics  and Astronomy,
University of Sussex
Brighton, BN1 9QH, UK \\
}

\date{Accepted XXX. Received YYY; in original form ZZZ}

\pubyear{2021}

\begin{document}
\label{firstpage}
\pagerange{\pageref{firstpage}--\pageref{lastpage}}
\maketitle

\begin{abstract}
Using data from the Pantheon SN Ia compilation and the Sloan Digital Sky Survey (SDSS), we propose an estimator for weak lensing convergence incorporating positional and photometric data of foreground galaxies. The correlation between this and the Hubble diagram residuals of the supernovae has $3.6\sigma$ significance, and is consistent with weak lensing magnification due to dark matter halos centered on galaxies. We additionally constrain the properties of the galactic haloes, such as the mass-to-light ratio $\Gamma$ and radial profile of the halo matter density $\rho(r)$. We derive a new relationship for the additional r.m.s. scatter in magnitudes caused by lensing, finding  $\sigma_{\rm lens} = (0.06 \pm 0.017) (d_{\rm C}(z)/ d_{\rm C}(z=1))^{3/2}$ where $d_{\rm C}(z)$ is the comoving distance to redshift $z$. Hence the scatter in apparent magnitudes due lensing will be of the same size as the intrinsic scatter of SN Ia by $z \sim 1.2$. We propose a modification of the distance modulus estimator for SN Ia to incorporate lensing, which can be easily calculated from observational data. We anticipate this will improve the accuracy of cosmological parameter estimation for high-redshift SN Ia data.
\end{abstract}





\section{Introduction}
\label{sec:intro}

Type Ia supernovae (SN Ia) are used extensively in cosmology as standard candles, due to an empirical relation between their absolute magnitudes and observable light curve properties \citep{Phillips1993, Tripp1999}. Their high luminosities allow them to be observed out to redshift $z \sim 2$ \citep{Riess2018c}. Thus they serve as a key cosmological resource, connecting the expansion history of the universe from when it was matter-dominated, through to the current epoch of dark energy domination. 
\par
The relative luminosities of SN Ia may be assembled in a Hubble diagram and used to constrain cosmological parameters such as the matter density $\Omega_{\rm M}$ and the equation of state of dark energy $w$ in simple extensions of $\mathrm{\Lambda CDM}$ \citep{Scolnic2018}. If their absolute magnitude is calibrated, the Hubble constant $H_0$ may be determined (for example see \citet{Riess2021, Freedman2019, Lemos2019}, and for a review of the distance ladder see \citet{Shah2021}).
\par
However, for supernovae to be accurate and unbiased distance estimators, the scatter of their observed magnitudes must be well-characterised. Gravitational lensing forms an important part of the scatter because it is an effect that increases with distance. Treated as random scatter, it therefore degrades the precision of survey data \citep{Holz2005}. Also, for magnitude-limited surveys, the calculation the Malmquist bias correction requires an understanding of the sources of scatter \citep{Kessler2017}. Therefore, to make the most of modern high-redshift SN Ia data sets, it is essential to understand lensing magnification. 
\par
Gravitational lensing can also be used as a cosmological probe in its own right. The lensing signal is sensitive to the amount, distribution and type of dark matter. For example, \citet{Metcalf1999} and \citet{Seljak1999} examined the case of dark matter being a mixture of weakly-interacting massive particles (WIMPs) or massive compact halo objects (MACHOs; for example, black holes). They showed that the skew of the lensing probability distribution function (pdf), proxied by the difference between the mode and the mean, is sensitive to $\Omega_{\rm m}$ and $\Omega_{\Lambda}$ independently, but most sensitive to the form of dark matter. They argued that even a modest sample of 100 SN Ia would be sufficient to constrain the fraction of MACHOs to within $20\%$. In a similar argument, \citet{Hada2016} show how SN Ia magnification may be used to bound the sum of neutrino masses. Going further, the moments of the lensing pdf may be fitted by simulations to the power spectrum of matter density fluctuations \citep{Marra2013}. This is particularily interesting in the context of moderate tensions that have arisen between measurements of the power spectrum normalisation $\sigma_8$ from the cosmic microwave background (CMB) and galaxy surveys (see for example \citet{Lemos2021, Troster2021}). 
\par
Gravitational lensing magnification is also complementary to time-delay and shear lensing studies, which have been used to measure the Hubble constant $H_0$ \citep{Wong2019}, and build maps of foreground mass \citep{Oguri2018, Giblin2021, Jeffrey2021}). However these studies have some drawbacks. They are distance-limited, as the shape of the galaxy must be resolved for it to work. Nuisance parameters and some bias may be introduced by an intrinsic alignment model. They are low-resolution, and require sufficient numbers to average over. Lastly, they suffer from the \say{mass-sheet degeneracy} whereby the cosmological parameters, but not the lensing observables, are changed by the addition of a matter sheet of constant density along the line of sight (see section 5.2 of \citet{Bartelmann2001}). A measurement of the absolute magnification of a background source breaks this degeneracy \citep{Falco1985}.
\par 
Type Ia supernovae seem ideal for magnification studies as, once standardized, their luminosities have an intrinsic scatter of $\sigma_{\rm int} \simeq 0.1$ mag. Secondly, SN Ia can be seen at distances great enough for the magnification to be measurable. The scatter caused by magnification is thought to be $\sigma_{\rm lens} = 0.04z - 0.09z$ mag where $z$ is the redshift, on the basis of simulations \citep{Frieman1997,Holz2005, Marra2013}. A small number of well magnified SN Ia with $\Delta m <-0.25$ mag are expected. However, four problems exist in observing it. Firstly, existing survey numbers peak at $z \sim 0.3$ where the magnification is likely to be small. Secondly, SN Ia are rare, transient events, so the sample to work with is smaller by two orders of magnitude compared to galaxy or quasar samples. Thirdly, the analysis must be centred around the set of sources, rather than lenses (as weak lensing surveys are), so it is \say{pot luck} what lies close to the line of sight (LOS). Additionally, the limiting magnitude (also known as the \textit{detection efficiency}) of a supernovae survey is not so straightforward to determine, meaning biases are harder to estimate. Lastly, it is far from certain that a limited set of SN Ia would fairly sample the distribution of magnification, and selection processes could obscure its effect (we expand on this further in Section 5). 
\par
Observational studies of SN Ia magnification have been made by some authors. \citet{Jonsson2010} used 175 SN Ia from the Supernova Legacy Survey (SNLS) to detect magnification at a confidence of $\sim 1.4 \sigma$, and estimated that lensing contributes an extra dispersion of $\sigma_{\rm lens} = 0.055 z$ magnitudes to supernovae. This value continues to be used in most cosmological analyses involving SN Ia, notably by the SH0ES team \citep{Riess2021} in estimating $H_0$, and is embedded in Pantheon SN Ia data \citep{Scolnic2018}. \citet{Smith2014} combined a larger sample of 608 SDSS SN Ia with number counts of a homogeneous sample of foreground galaxies. The authors also found a detection significance of $\sim 1.4 \sigma$. \citet{Macaulay2020} used SN Ia and galaxies from the Dark Energy Survey (DES) 1Y data to compute the skew of the magnitude distribution (assuming intrinsic scatter is Gaussian, the skew may be attributed to lensing) with a simulated fit given by \citet{Marra2013} to derive a constraint on the matter-power spectrum, and report a $\sim 1.3 \sigma$ detection of magnification. 
\par
In this paper, we have two main goals. Firstly, we seek to establish lensing is occurring with more certainty and measure its scatter. Secondly, we aim to constrain the mass and profile of galactic dark matter haloes within our chosen modelling framework. Adopting a Bayesian approach, we test a two-parameter family of physically-motivated halo profiles, including the Navarro-Frenk-White (NFW, \citet{Navarro1996}) and Hernquist \citep{LarsHernquist1990} profiles as special cases. We calculate posterior distributions of the parameters that describe our haloes.
\par
Our paper is organised as follows. In Section \ref{sec:WLbyhalos}, we derive the magnification in the weak-lensing regime of our halo profile, and connect this to lensing over cosmological distances. In Section \ref{sec:data}, we describe our data and selection criteria. In Section \ref{sec:method}, we specify our estimator for lensing convergence, what observables we will correlate, and our Bayesian model. Results are presented in Section \ref{sec:results}, which we compare to the literature. We summarise our results in Section \ref{sec:summary}. A future paper will be devoted to cosmological parameters derived using SN Ia lensing. We retain factors of $c$ in equations.

\section{Lensing model}
\label{sec:WLbyhalos}
\subsection{Weak lensing by halos}
For extended matter distributions it is useful to adopt the lensing potential formalism of \citet{Schneider1985}, for which we state the relevant formulae in this subsection. The \textit{convergence} is defined as 
\begin{equation}
\label{eq:conv}
    \kappa = \frac{\Sigma(\vec{\theta})}{\Sigma_{\rm c}} \; ,
\end{equation}
where the \textit{surface density} is
\begin{equation}
\label{eq:surfdens}
    \Sigma(\vec{\theta}) = \int \rho(\vec{\theta}, z) dz \;,
\end{equation}
and the \textit{critical surface density} is
\begin{equation}
\label{eq:critsurfdens}
    \Sigma_{\rm c} = \frac{D_{\rm s}}{D_{\rm d} D_{\rm ds}} \frac{c^2}{4\pi G} \;\;.
\end{equation}
Here $D_{\rm d}, \, D_{\rm s}, \, D_{\rm ds}$ are the angular diameter distances to the lens, source and between lens and source respectively. 
\par
We define $\delta m = m_{\rm lens} - m_{0}$ as the change in magnitude of a source due to a single lens, relative to the background cosmology in the absence of the lens. When the magnification is small, it may taken to first order in the convergence and shear $\gamma$ as 
\begin{align}
\label{eq:dmlens}
    \delta m  = -(5/\log{10}) \kappa + O(\kappa^2, \gamma^2) \;. 
\end{align}
Hence we can compute the magnification expected from a galactic halo of a given density profile $\rho(r)$ by the following recipe : 
\begin{itemize}
    \item Obtain the critical density $\Sigma_{\rm c}$ from the angular diameter distances $D_{\rm ds}, D_{\rm s}$ and $D_{\rm d}$ in the background cosmology by Equation (\ref{eq:critsurfdens}),
    \item Calculate the surface density $\Sigma$ from Equation (\ref{eq:surfdens}) and convergence $\kappa$ from Equation (\ref{eq:conv}). 
    \item Use Equation (\ref{eq:dmlens}) to calculate the magnification.
\end{itemize}
We note two important points. Firstly, magnification is a concave function of the distance from the observer to the lens, hence when magnification is strongest, it will be relatively insensitive to moderate errors in the distances (the gradient of magnification with lens distance will be small); this will be helpful when using distances derived from photometric redshifts. Secondly, as magnification is inversely proportional to (a power of) the impact parameter $b$ via Equation (\ref{eq:surfdens}), its distribution will be skewed towards high magnification. Therefore, the commonly-used assumption of Gaussian scatter in SN Ia residuals is not correct. 
\par 
Equation (\ref{eq:dmlens}) is valid to first order in the gravitational potential $\Phi/c^2$, the light deflection angle $\alpha$, and the convergence $\kappa$. These assumptions can be expected to hold for lensing by extended diffuse halos; \citet{Jonsson2010} has checked the validity of this against ray-tracing around such halo types, finding the difference to be less than $5\%$. We have also checked including shear for the NFW halo does not change our results.

\subsection{Halo model}
We define a double-power law halo with profile
\begin{equation}
    \label{eq:doublepowerlawhalo}
    \rho (r; \gamma, \beta) = \frac{\delta_{\rm c} \rho_{\rm c}}{(\frac{r}{r_{\rm s}})^{\gamma} (1+(\frac{r}{r_{\rm s}}))^{\beta}} \; .
\end{equation}
where $\rho_{\rm c} = 3 H(z)^2 / 8\pi G$ is the critical density of the universe at redshift $z$, $\delta_{\rm c}$ is a density parameter and $r_{\rm s}$ is the \textit{scale radius}. Our data will not have sufficient resolving power to constrain the inner slope $\gamma$, so we fix $\gamma = 1$ and leave $\beta$ as a free parameter. Then, at small radii $\rho \; \propto \; r^{-1}$, whereas at large radii $\rho \;\propto \; r^{-1 - \beta}$.
\par 
Standard spherical collapse theory implies that a sphere of radius $r_{200}$ with average density $\bar{\rho} = 200 \rho_{\rm c}$ may be considered gravitationally bound. The scale radius $r_{\rm s}$ is then defined relative to $r_{200}$ as
\begin{equation}
    r_{\rm s} = r_{200} / c
\end{equation}
where $c \simeq 5 - 15$ is the \textit{concentration}\footnote{Any potential confusion with the speed of light $c$ should be clear from the context.} which we take as a second free parameter. We make $r_{200}$ a function of the mass $M_{200} = M(r<r_{200})$ it encloses
\begin{equation}
\label{eq:m200}
    M_{200} = \frac{800\pi}{3} \rho_{\rm c} r_{200}^{3} \; ,
\end{equation}
from which we obtain $\delta_{\rm c} = \frac{200}{3} c^3 f(c)$ where 
\begin{equation}
    f(c) = 
    \begin{cases}
        \frac{1}{c - \ln{(c+1)}} & \beta=1 \\[1em]
        \frac{1}{\ln{(1+c)} - c/(1+c)} & \beta = 2 \\[1em]
        \frac{ (\beta - 2)(\beta - 1)}{(1-(c+1)^{1-\beta} ((\beta -1) c +1))} & \beta \neq 1,2 \;\; .\\
    \end{cases}
\end{equation}
Note that $\beta = 2$ corresponds to the NFW profile \citep{Navarro1996}, and $\beta = 3$ is the Hernquist profile \citep{LarsHernquist1990}. The convergence and shear have analytical formulae for integer $\beta$, and we state those formulae (which have been derived in the literature) together with that for a singular isothermal sphere, in Appendix \ref{sec:haloprofile}. 
\par
\begin{figure}
    \centering
    \includegraphics[width=\columnwidth]{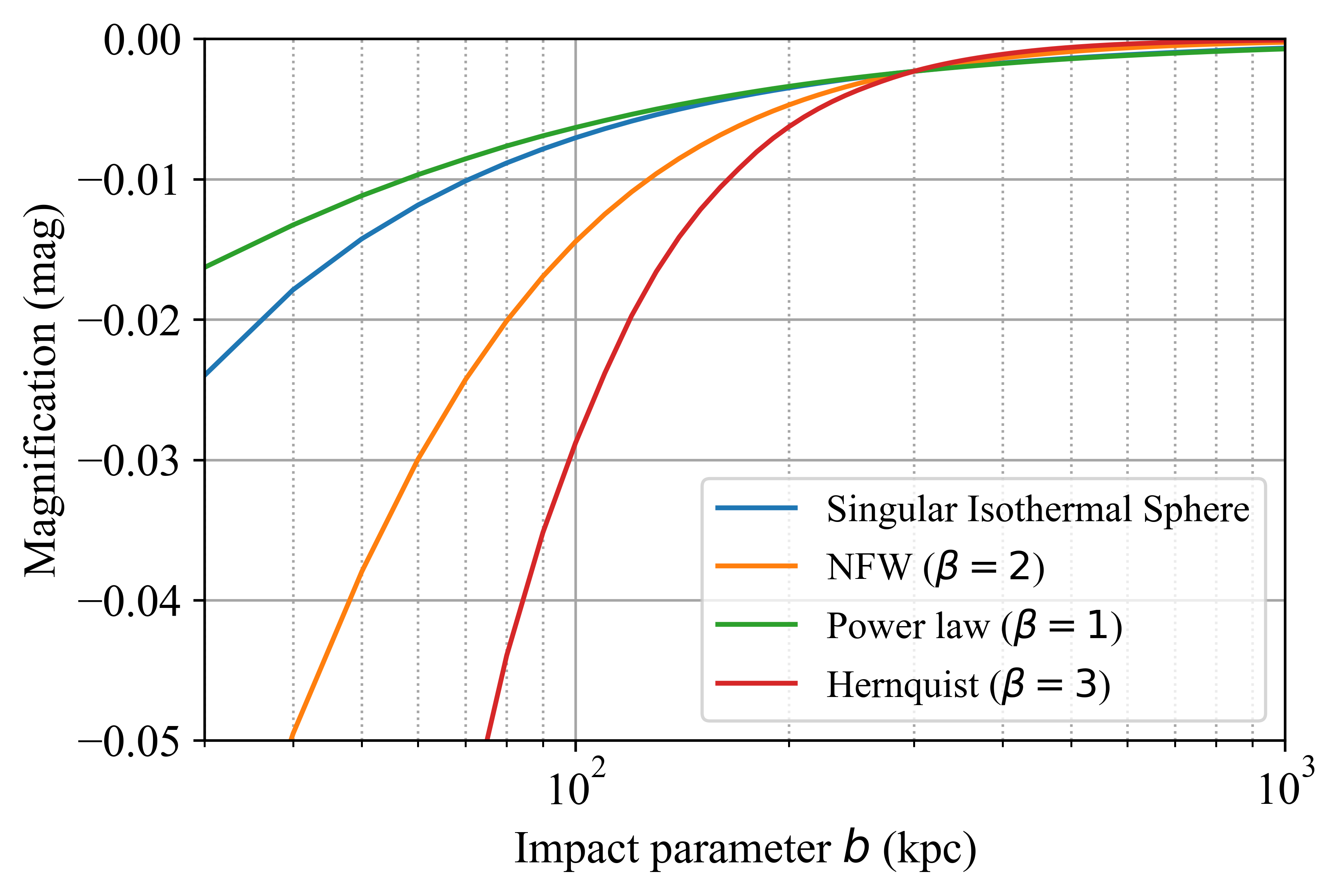}
    \caption{The magnification $\delta m$ (in units of magnitude) by a typical galaxy of absolute magnitude $M = -21.25$ located at $z = 0.2$, of a source located at $z = 0.5$. The mass of each halo profile has been normalised to give the same observed magnification at $b = 300$ kpc, about the magnification-weighted average of our data sample. Although SN Ia lines of sight passing within 100 kpc of a lensing galaxy centre are rare, we obtain the most constraining power from them.}
    \label{fig:halobetaprofile}
\end{figure}

 
For general $\beta$, we must obtain the surface density numerically. This is simplified by making the substitution $t = \sqrt{ r/(r+r_{\rm s}) (x+1) - x}$ which results in 
\begin{equation}
\label{eq:betasurfdens}
    \Sigma(x ; \beta) = 4 \delta_{\rm c} \rho_{\rm c} r_{\rm s} (x+1)^{\frac{1}{2} - \beta} \int_{0}^{1} \frac{(1-t^2)^{\beta - 1}}{\sqrt{(1-x)t^2 +2x}} dt\;\;,
\end{equation}
where 
\begin{equation}
    x = b / r_{\rm s}
\end{equation}
is the dimensionless impact parameter in units of the scale radius. We illustrate the magnification calculation for a range of our halo profiles in Figure \ref{fig:halobetaprofile}.

\par
We may expect many galaxies to lie at a large distances from the LOS, so it is worthwhile to examine the lensing profile at large $x$. Taking the limit of Equation (\ref{eq:betasurfdens}) and re-expressing the surface density as a function of the impact parameter $b$ we find 
\begin{equation}
   \Sigma(x, \beta) \longrightarrow D \times (M_{200})^{\frac{\beta+1}{3}} \frac{f(c)}{c^{\beta-2}}  \rho_{c}^{\frac{2-\beta}{3}} \;\; \frac{1}{b^{\beta}} \qquad \mbox{for $x \gg1$},
\end{equation}
where $D$ is a numeric constant of order unity. We see from the above that, as expected $\Sigma \; \propto \; b^{-\beta}$, but in the pre-factors there is a degeneracy between the mass of the halo $M_{200}$ and a function of the concentration parameter $c$ in the large $x$ limit.
\par 
Hence, for a given fixed slope $\beta$, a heavy but high concentration halo will magnify to the same extent as a lighter, less concentrated halo. We will therefore find it useful to adopt a model for $c$.

\begin{figure}
    \centering
    \includegraphics[width=\columnwidth]{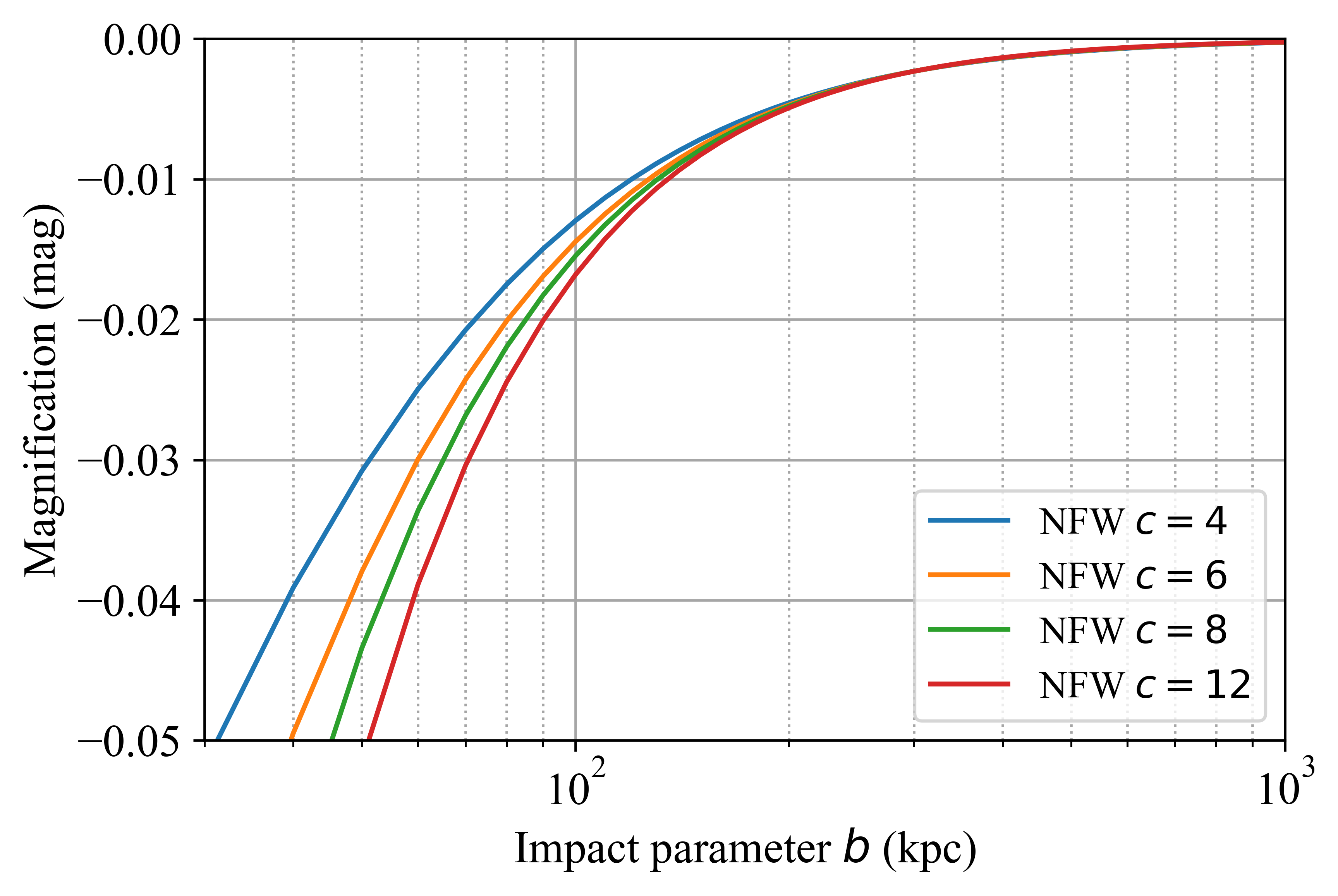}
    \caption{The magnification (in units of magnitude) due to a typical galaxy as detailed in Figure \ref{fig:halobetaprofile}. Here we illustrate the NFW profile magnification at different values of the concentration parameter $c$, again normalising to the same observed magnification at $b = 300$ kpc. It is evident from the graph that $c$ will only be weakly constrained by our data. }
    \label{fig:halocprofile}
\end{figure}

\subsubsection{Concentration models}
In the original paper of \citet{Navarro1996}, the NFW halo was characterised as a one-parameter family where $c \equiv c(M_{200})$. Recent studies have updated this result. \citet{Duffy2008} (hereafter D08) used WMAP-derived cosmological parameters to simulate halos in the mass range $10^{11} - 10^{15} h^{-1} M_{\odot}$, fitting 
\begin{equation}
    c = A (M_{200} / M_{\star})^B (1+z)^C \;,
\end{equation}
where $A = 5.71 \pm 0.12$, $B = -0.084 \pm 0.006$ and $C = -0.47 \pm 0.04$. \citet{Munoz-Cuartas2011} (hereafter MC11) also use a WMAP-derived cosmology with similar resolution and fit 
\begin{align}
    \log{c} & =  a(z) \log{M_{200}/[h^{-1}M_{\odot}]} + b(z) \\
    a(z) & =  wz - m \\
    b(z) & =  \frac{\alpha}{(z+\gamma)} + \frac{\beta}{(z+\gamma)^2} \;\;,
\end{align}
where $w = 0.029$, $m = 0.097$, $\alpha = -110.001$, $\beta = 2469.72$, and $\gamma = 16.885$. Finally, \citet{Mandelbaum2008} (hereafter M08) compile weak lensing analyses of SDSS galaxies in the mass range $10^{12} - 10^{14} h^{-1} M_{\odot}$ and fit
\begin{equation}
    c = \frac{c_0}{(1+z)} \left( \frac{M}{M_{\star}} \right)^{-\beta} \;,
\end{equation}
where $c_0 = 4.6 \pm 0.7$, $\beta = 0.13 \pm 0.07$ and $M_{\star} = 1.56 \pm 0.12 \times 10^{14} h^{-1} M_{\odot}$. 
\par 
For our data, we find $\langle c \rangle = 5.2, 8.0, 6.9$ for the D08, MC11, M08 models respectively. We test our results on all concentration models, and also $c$ as a uniform global parameter for comparative purposes. For our main result we will prefer the M08 model as it has been derived from observations.

\subsection{Homogeneous cosmology}
For our background cosmology, we assume a spatially-flat $\Lambda$CDM model, and neglect $\Omega_r$ so that $\Omega_\Lambda = 1 - \Omega_{\rm m}$. We have for the angular diameter distance $D_{\rm A}$ and luminosity distance $D_{\rm L}$ the standard formulae
\begin{align}
\label{eq:cosmoformulae}
    D_{\rm L} (z) & =  \frac{c}{H_0} (1+z_{\rm obs}) \int_{0}^{z_{\rm cos}} \frac{dz'}{E(z')} \; , \\
    D_{\rm A} (z) & = D_{\rm L} / (1+z_{\rm obs})^2, \nonumber \\
    E(z) & = \sqrt{\Omega_{\rm m,0} (1+z_{\rm cos})^3 + \Omega_\Lambda,0} \; , \nonumber
\end{align}
where $H_0$ is the present day Hubble constant, $H(z) = H_0 E(z)$ and  $\Omega_{\rm i,0}$ is the present day component density. $z_{\rm obs}$ refers to the observed heliocentric redshift, and $z_{\rm cos}$ the redshift corrected for peculiar velocity to the CMB rest-frame. When using standard candles, it is convenient to re-express the luminosity distance as the distance modulus
\begin{equation}
    \label{eq:mumodel}
    \mu(z) = 5 \, \mbox{log}_{10} (D_{\rm L}(z)/10 \mbox{pc}) \; .
\end{equation}
Our results depend only very weakly on the cosmological parameters used (via the angular diameter distances used in $\Sigma_{\rm crit}$, and $\rho_{\rm c}$ used to normalise the halo density), except in the case of the physical mass-to-light ratio. However, to be concrete we set $h = 0.674$ and $\Omega_{\rm m} = 0.298$, where $H_0 = 100 h \;\SI{}{\kilo\meter\per\second\per\mega\pc}$ and $\Omega_{\rm m}$ is the best fit to the Pantheon sample.

\subsection{Density model}
Our model for the matter density is 
\begin{equation}
\label{eq:rhomodel}
        \rho(\vec{r},z) = \rho_{\rm void}(z) + \sum \rho_{\rm halo}(\vec{r}, \vec{r}_i,z)
\end{equation}
where $\rho_{\rm halo} (\vec{r},\vec{r}_i,z)$ is the density profile of a dark matter halo located at $\vec{r}_i$ and redshift $z$. $\rho_{\rm void}(z)$ is a spatially uniform minimum density that is a function of redshift only; it represents the average remaining density of the universe if the virial masses of galactic halos were removed, and is determined by the requirement that $\bar{\rho} = \rho_{\rm c}$. We take the form of these halos to be the spherically-symmetric profiles as described in the previous section. Although we can in general expect the halos to be non-spherical, it has been shown that after taking the average over randomly oriented non-spherical halos for a lensing calculation, spherical-symmetry is a very good approximation \citep{Mandelbaum2005}. We neglect additional inhomogeneous contributions due to filaments or sheets, and assume baryons are distributed with the same profile as dark matter for the purposes of a magnification calculation. 

\subsubsection{Flux conservation}

It can be shown that in the weak lensing approximation, the average magnification (over a large number of sources) compared to a homogeneous background is unity (see for example \citet{Kainulainen2009}). In fact this argument, which was originally made in a more general context by \citet{Weinberg1976}, depends on three key assumptions. Firstly, the universe is assumed to be transparent and lines of sight are not \say{special} in some way (Weinberg argued that if galaxies are opaque disks, and hence the lines of sight are those that are unobscured by foreground galaxies, the average result will be de-magnification; \citet{Kainulainen2011} give a quantitative prescription for calculating this effect in terms of a survival probability as a function of impact parameter). Secondly, the distance of sources is unaffected by lensing, which introduces perturbations to the sphere of constant redshift. \citet{Kaiser2016} have shown this is equivalent to working to first order in the convergence $\kappa$. Thirdly, by working to first order in post-Newtonian potential $\Phi$, the \say{back reaction} of inhomogeneity on spacetime is neglected and it is assumed that the homogeneous universe formulae (\ref{eq:cosmoformulae}) may continue to be used. What may happen in the more general non-perturbative case is still the subject of active research (see for example \citet{Buchert2015}).
\par
Therefore, working to first order in the convergence\footnote{It is the flux of photons that is conserved on average, not any non-linear quantity such as magnitude derived from it. However the correction is second order and we neglect it.}, we take
\begin{equation}
\label{eq:fluxcons}
    \langle \Sigma_j \delta m_{i,j} \rangle = 0 \;\;,
\end{equation}
where sum is over $j = 1 \ldots N_i$ foreground galaxies and the average is over $i = 1 \ldots N_k$ sources in the redshift bin $z \in (z_k, z_{k+1})$. For lensing by galactic halos at $z<1$, our typical calculated $\Delta m \equiv \Sigma_j \delta m$ are $\mathcal{O} (10^{-2})$ to $\mathcal{O} (10^{-3}) $. We therefore expect Equation (\ref{eq:fluxcons}) to be a good approximation at redshifts $z<1$ for sources that are weakly lensed. However, were we to be analysing sources at redshift $z \sim 2$ with $\Delta m \sim \mathcal{O} (10^{-1})$, second order effects should not be ignored.

\section{Data}
\label{sec:data}

\subsection{Supernovae}
SN Ia magnitudes are standardised by the Tripp estimator \citep{Tripp1999}, which is a function of observable features of their light curves. A commonly-used form expresses the distance moduli $\mu$ of a SN Ia as 
\begin{equation}
\label{eq:tripp}
\mu = m_B - M_B + \alpha x_1 - \beta c + \Delta_{\rm M} + \Delta_{\rm B} \; ,
\end{equation}
where the observables are $m_B$, the peak apparent AB magnitude of the supernova; $x_1$, the \say{stretch} of its light curve (a dimensionless parameter typically between -2 and 2 representing the duration of the curve); and $c$, the deviation of the B-V colour from the mean colour. $M_B$ is the absolute magnitude of a fiducial mean SN Ia light curve. $\Delta_{\rm M}$ is a correction based on the host galaxy mass or other environmental effects, and $\Delta_{\rm B}$ is a bias correction derived from simulations to account for the selection process of the sample. The nuisance parameters $\alpha$ and $\beta$ are fitted for to minimize residuals versus a background cosmology.
\par
For this analysis, we use the Pantheon data set\footnote{\url{https://github.com/dscolnic/Pantheon}} \citep{Scolnic2018}. The file \path{ANCILLIARY_G10.FITRES} contains the apparent magnitudes (adjusted using Eqn.~\ref{eq:tripp}), redshifts, positions, stretch $x_1$ and colour $c$ of 1,048 SN Ia compiled from multiple surveys, with bias corrections determined according to the intrinsic scatter model of \citet{Guy2010}. We also use the covariance matrix \path{sys_full_long.txt}. We select the 901 SN Ia within the SDSS footprint.

\subsubsection{Field selection}
To obtain a clean lensing signal, we must identify the host galaxy : the supernovae will not be lensed by its host halo. If we fail to identify the host, due to errors in galaxy redshifts (see below) the true host may be present in the foreground close to the line of sight, and so add a large spurious amount to the lensing estimate. We identify the galaxy closest in angular distance to the location of the supernova as the host, and exclude it from our lensing estimator. We also check that the redshift of the putative host is compatible with the redshift of the SN Ia. A limitation of this is that the SDSS photometric redshift confidence intervals are not always reliable, so in our results section we also quote the correlation using angular distance only for comparison. 
\par 
To check we have indeed correctly identified the host, we also calculate the impact parameter of the closest galaxy in units of galactic scale length $x = b / r_{\rm s}$. Large values will indicate doubt that we have identified the correct host, and we discard fields with $x>4$ (corresponding to 100 kpc for a typical galaxy). The cutoff for $x$ should not be too small as to unnecessarily remove supernovae originating in faint nearby satellite galaxies or the stellar halo, but not extend past the virial radius $x \sim c$. This selection reduces the sample to 762 supernovae.
\par
To reduce noise, we further exclude fields that are more than $50\%$ masked in the foreground galaxy sample, as the lensing estimate is unlikely to be accurate for them. $35\%$ of our fields have some masking in them, with the average fraction of masked objects in those fields $\sim 20\%$. Our final sample therefore comprises 720 SN Ia. 

\subsection{Galaxies}
We use galaxies drawn from the SDSS \citep{Eisenstein2011}. The survey performed deep imaging of 8400 deg$^2$ of the high Galactic latitude sky, and spectroscopy of over 1.5 million galaxies. A supernovae survey was conducted of the so-called \say{Stripe 82}, a strip $2.5^{\circ}$ wide along the celestial equator from right ascension $20^h$ to $4^h$, and this is a prominent overlap with the Pantheon footprint. The galaxy survey is expected to be 95\% complete at $r = 22.2$ magnitude limit \citep{Abazajian2009}, and the faintest sources categorised as galaxies are up to $r \sim 26$.
\par
We initially select all photometric objects in the SDSS Data Release 16 view \path{Galaxy} which are in an aperture of radius 8 arcminutes around the line of sight to each supernova. For redshift $z=0.2$, this corresponds to a physical distance of $\sim 1.5$ Mpc. We discuss and test the choice of aperture in Section \ref{sec:results}. We find 728,280 galaxies, of which 176,872 are in the foreground of their supernovae. The average redshift of our foreground sample is $z \sim 0.34$. We select photometric data in the $gri$ passbands, using the \texttt{cModelMag} magnitudes $m_{\lambda}$ as the best representation of the brightness of galaxies.
\par
\begin{figure}
    \centering
    \includegraphics[width=\columnwidth]{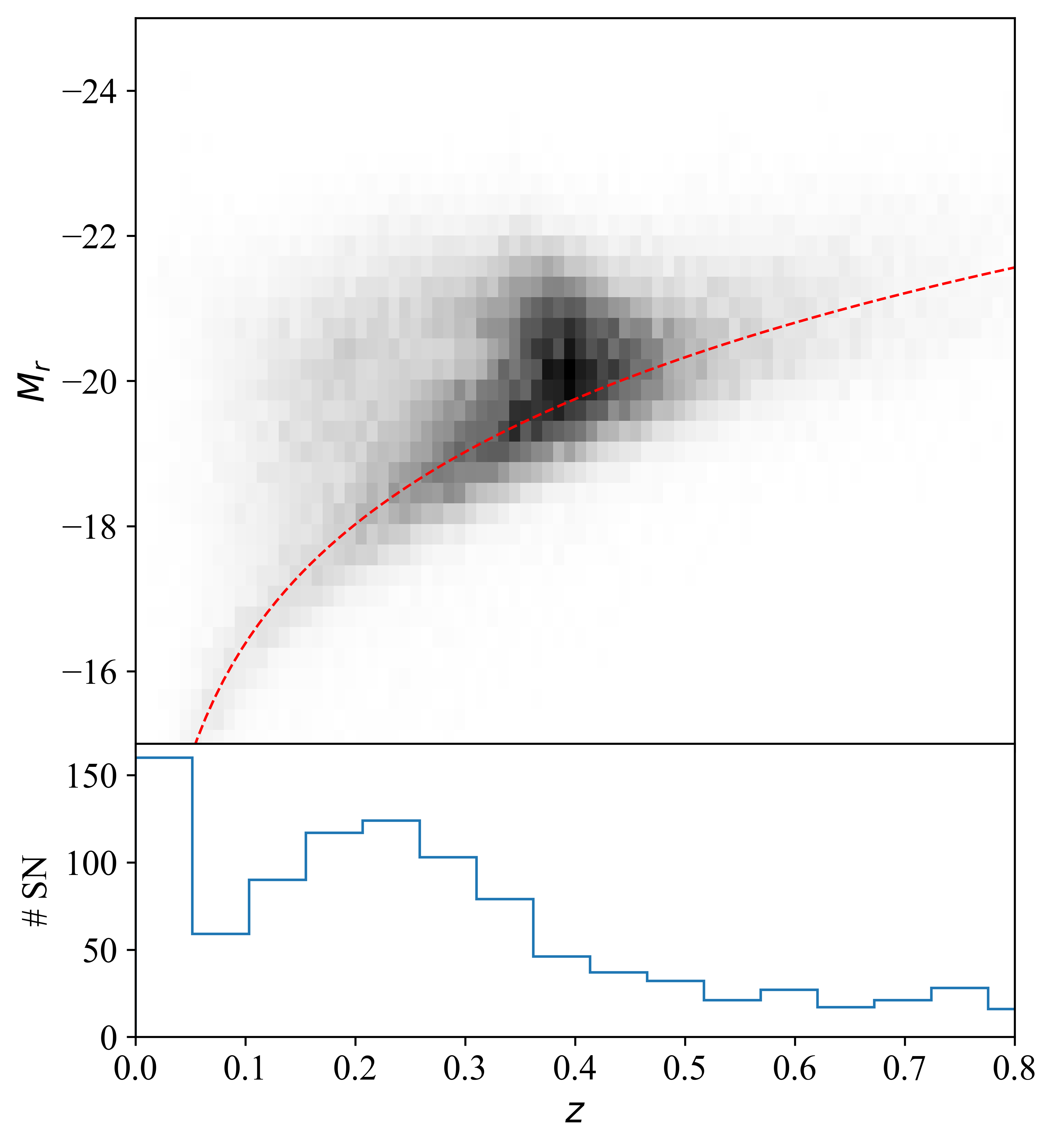}
    \caption{Number distribution of our galaxy and supernovae sample by calculated r-band absolute magnitude $M_r$ and redshift $z$. The red dotted line shows the SDSS survey limiting magnitude $m_r = 22$, the effect of which is clear in the sample.}
    \label{fig:galaxyandsndist}
\end{figure}

 
We select galaxies with clean photometry identified by the flags \texttt{InsideMask}$=0$ and \texttt{Clean}$=1$, which reduces our sample size by $33\%$. We derive the absolute magnitude $M_{\lambda}$ of the galaxy in a given passband as 
\begin{equation}
    M_{\lambda} = m_{\lambda} - \mu(z_p) - K_{\lambda} - A_{\lambda} \;,
\end{equation}
where the survey-reported K-corrections $K_{\lambda}$ and Milky Way extinction $A_{\lambda}$ are used. The distance modulus $\mu$ is derived using the photometric redshift $z_p$ by Equations (\ref{eq:cosmoformulae}, \ref{eq:mumodel}). We floor $M_{\lambda}$ at $-25$ to reduce outliers. Our sample is illustrated in Figure \ref{fig:galaxyandsndist}.

\subsubsection{Photometric redshifts}
We use the SDSS photometric redshifts $z_p$ that were last updated in Data Release 12. The methodology used to determine them is outlined by \citet{Beck2016}. The errors are calculated for each galaxy and are distributed as $\delta z_p \sim 0.09 \pm 0.04$, and the relative error $\delta z_p / (1+z_p)$ is largely uncorrelated with redshift. The algorithm provides a flag \texttt{photoErrorClass} to indicate the quality of the fit, which is given values between $-7$ and $+7$. There is some correlation between \texttt{photoErrorClass} and redshift, and to minimize selection bias we therefore only exclude galaxies with \texttt{photoErrorClass}$<-4$, which is indicative of a poor-quality extrapolation from the training set. 
\par
For each galaxy $j$ in the field of SN Ia $i$ we compute the angular diameter distances and impact parameter $b = \theta D_d$ using formulae (\ref{eq:cosmoformulae}), using the photometric galactic redshift $z_p$ and the spectroscopic SN Ia redshift $z$. We also use $z_p$ to determine the average matter density at the location of the galaxy $\rho_{\rm c}(z) = 3 H(z)^2 / 8 \pi G$ which given the mass of the halo $M_{200}$ will determine its physical radius $r_{200}$ by Equation (\ref{eq:m200}).
\par 
We have already discussed that the lensing efficiency (Equation \ref{eq:critsurfdens}) is relatively insensitive to redshift error near its peak. Whilst the average relative redshift error on the whole galaxy sample is $\sim 35\%$, we find a relative error on the lensing efficiency of $\sim 20\%$. 
\par
Our final sample then comprises 68,039 galaxies. Our selection parameters are summarised in Table \ref{tab:parameters}, and in Section \ref{sec:results} we test the dependence of our results on these choices. 

\begin{table*}
    \centering
    \begin{tabular}{l l l}
    Parameter & Rationale & Value \\ \hline
    Max radius & Include galaxies within an annulus (arcminutes) & 8 \\
    \texttt{photoErrorClass} & Exclude galaxies with low quality redshifts & -3 to +7 \\
    \texttt{InsideMask} & Exclude galaxies that are in masked areas & 0 \\
    \texttt{Clean} & Exclude galaxies with poor photometry & 1 \\
    Magnitude & Cap abs mag of galaxies to reduce outliers & -25 \\
    Concentration & Cap derived concentration to reduce outliers & 15 \\ \hline
    Masked fraction & Exclude SN Ia with heavily masked foregrounds & 50\% \\
    Host radius & Exclude hostless SN Ia (within scale radii) & 4  \\
    Concentration model & Degeneracy between halo mass and concentration & M08  \\         &  \\
         & 
    \end{tabular}
    \caption{Summary of parameter choices and selection criteria of foreground galaxies. See Section 2.5 for a description of the concentration models. }
    \label{tab:parameters}
\end{table*}

\section{Methodology}
\label{sec:method}
\subsection{Estimating the lensing signal}
We take dark matter haloes to be aligned to the photometric centre of individual galaxies. Hence for magnification of supernovae $i$ due to galaxy $j$ we set 
\begin{equation}
    \delta m_{i,j} \equiv \delta m(b_{ij}, \mathbf{\varpi}_j)
\end{equation}
where $\delta m$ is given by Equation (\ref{eq:dmlens}). The impact parameter is $b_{ij} = \theta_{ij} D_{\textrm{d}, j}$ where $\theta_{ij}$ is the angular separation between galaxy $j$ and supernova $i$ and $D_{\textrm{d}, j}$ the angular diameter distance to the galaxy.
\par 
The halo parameters are $\mathbf{\varpi}_j =  \{ M_{200, j}, c_j, \beta \}$. To estimate $M_{200}$ we convert the galactic absolute magnitude $M_{\rm \lambda}$ using a halo mass-to-light ratio $\Gamma$ by 
\begin{equation}
    \label{eq:galaxymass}
    M_{200} = \Gamma(M_{200},p) 10^{0.4(M_{ \odot, \lambda} - M_{\lambda})}
\end{equation}
where $M_{ \odot, \lambda}$ is the solar absolute magnitude. $\Gamma (M_{200},p)$ will in general depend on both $M_{200}$ and galaxy type $p$ (for example morphology or colour).
\par 
Previous weak lensing shear studies have examined the relationship between mass-to-light ratios and luminosities, morphological type or colour. \citet{Mandelbaum2006} derive a relation between the luminosity of \textit{central} galaxy in a cluster and the \textit{total} cluster mass of $M \propto L^2$, but this is not applicable for our model. \citet{VanUitert2011} finds little dependence of $\Gamma$ in the SDSS sample on luminosity for $L < 7 \times 10^{11} L_{\odot}$, but early-type galaxies are heavier (see Table 2 and Figure 9 of that paper).  \citet{Brimioulle2013} finds $\Gamma \;\propto \;L^{0.12 \pm 0.11}$ for galaxies from the Canada-France-Hawaii Telescope Legacy Survey (CFHTLS), with red (defined as $B-V >0.7$) galaxies being heavier than blue at the same luminosity.
\par 
We do not have morphological information $p$ for most galaxies in our sample. Additionally, the size of our SN Ia sample is not sufficient to adequately constrain mass-to-light ratios for subsamples by colour or luminosity. Therefore we adopt $\Gamma (M_{200},p) \equiv \Gamma$ as a uniform sample average for our analysis. 
\par 
Summing contributions of individual galaxies we obtain the unnormalised magnification 
\begin{equation}
\label{eq:unnormmag}
    \Delta m^{\prime}_{i} = \sum_{j = 1}^{N_i} \delta m_{i,j} \;,
\end{equation}
where $N_i$ is the number of foreground galaxies in the supernova field.
\par
We impose the flux conservation of Equation (\ref{eq:fluxcons}) and define our \textit{magnification estimator} as
\begin{equation}
\label{eq:normmag}
    \Delta m_i = \Delta m_i^{'} - \langle \Delta m_k^{'} (z_i) \rangle  \;.
\end{equation}
In the second term, the average is taken over all supernovae $i$ in the redshift bin $z_k < z_i < z_{k+1}$. Hence, by construction $ \langle \Delta m_i \rangle= 0$ in each bin (although our result for correlation doesn't depend on this). In practice, most lines of sight do not pass very close to a foreground galaxy, and those supernovae will be mildly de-magnified. A smaller number will be magnified and hence the distribution of $\Delta m_i$ is skewed with the median and mode positive.

\subsubsection{SN Ia colour}
The SALT2 light-curve fitter for supernovae \citep{Guy2007} outputs a colour parameter $c = (B-V) - \langle B-V \rangle$ which is the difference between the colour at peak B-band magnitude and the average for the training sample\footnote{The distinction between the NFW concentration parameter $c$ and colour should be clear from the context.}. This may be interpreted as the sum of some intrinsic colour scatter, plus reddening E(B-V). Some portion of this reddening will be due to the host galaxy, and some will be due to extinction by dust embedded in foreground galactic halos. SN Ia magnitudes are de-reddened by the Tripp estimator (Equation \ref{eq:tripp}), which subtracts the reddening $\beta c$ where $\beta \sim 3$ is a fitted parameter consistent with $R_V = 3.1$.
\par
However, we may estimate the amount of reddening due to dust in foreground galactic haloes in the following way. The combined effect of magnification and extinction is
\begin{equation}
    F = F_0 \nu e^{- \tau_{\lambda}} \;, 
\end{equation}
where  $\nu = F_{\rm lens}/F_0$ is the lensing magnification factor, and $\tau_{\lambda}$ is the wavelength-dependent optical depth. In magnitudes, we obtain
\begin{equation}
    \Delta m (\lambda) \simeq 1.08(\tau_{\lambda} - \Delta \nu) \;,
\end{equation}
where $\Delta \nu = \nu - 1$. As lensing is achromatic, it follows that $E(B-V) \simeq 1.08 ( \Delta m (B) - \Delta m (V))$. \citet{Menard2010a} investigated dust extinction in galactic halos by correlating the colours and magnitudes of quasars between $1 < z < 2.5$ with the angular distance to SDSS foreground galaxies, finding the visual band magnification is offset about $1/3$ by extinction. 
\par 
By using the dust-to-mass ratio $\Upsilon = 1.1 \times 10^{-5}$ derived in \citet{Menard2010} and a typical host galaxy dust mass opacity $\kappa_V = 1540 \, \SI{}{\meter\squared\per\kilo\gram}$ \citep{Weingartner2001}, we estimate
\begin{equation}
\label{eq:deltaci}
    \Delta c \simeq 0.01694 \;\Sigma(r) \; \mbox{mag},
\end{equation}
where $\Sigma(r)$ is the surface density of the galactic halo. 

\subsubsection{SN Ia stretch}
Supernovae have a finite size, and there will be some time delay (due to both the differential path length and time dilation) between light arriving from opposite sides of the expanding photosphere. In principle at least, this may result in some of the magnification being subtracted out as an increased stretch parameter $x_1$ of the light curve when the magnitude is standardised by the Tripp estimator (see Equation (\ref{eq:tripp}) below). The differential time delay is proportional to the product of light travel time across the ejecta, and the deflection angle $\hat{\alpha}$ (which is the gradient of the time delay across the ray bundle). Thus, taking an ejecta velocity of $\sim 0.05 c$, a lightcurve duration of $\sim 30$ days, and deflection angle $10^{-2}$ rad, the differential time delay will be roughly 30 minutes. However, this translates into a change in stretch parameter of only $\Delta x \sim 5 \times 10^{-3}$ (for the relation between $x_1$ and the duration of the light curve, see Figure 2 of \citet{Guy2007}). Hence, the magnitude will be adjusted upwards by $\sim 0.001$ mag for a typical value of $\alpha \sim 0.15$ in the Tripp estimator. This is less than $5\%$ of the calculated magnification for the same light deflection angle. We therefore ignore this effect.  

\subsection{Correlating with Hubble diagram residuals}

We determined the SN Ia distance modulus residuals $\mu_{\rm res}$ by fitting a Hubble diagram to minimize 
\begin{equation}
    \chi^2 = \mathbf{\mu}_{\rm res}^{T} \cdot \mathbf{C^{-1}} \cdot \mathbf{\mu}_{\rm res} \;\;,
\end{equation}
where $\mathbf{\mu_{res}} = \mathbf{\mu} - \mathbf{\mu}_{\rm model}$ and $\mu_{\rm model}(H_0, \Omega_{\rm m}, z)$ is given by Equation (\ref{eq:mumodel}) where $\mu$ is the apparent SN Ia distance modulus $\mu = m - M$.  
\par
$\mathbf{C}$ is the Pantheon covariance matrix \citep{Scolnic2018}, which is the sum of statistical and systematic errors $\mathbf{C} = \mathbf{C}^{\rm stat} + \mathbf{C}^{\rm sys}$. The statistical error matrix is diagonal with entries given by 
\begin{equation}
    C^{\rm stat}_{ii} = \sigma_{N}^{2} + \sigma_{\rm Mass}^{2}+ \sigma_{\rm v-z}^{2}+ \sigma_{\rm lens}^{2} + \sigma_{\rm int}^{2} + \sigma_{\rm Bias}^{2} \;,
\end{equation}
where the largest term is the intrinsic SN Ia dispersion $\sigma_{\rm int} \sim 0.08$. The (Gaussian) term for the dispersion caused by lensing is $\sigma_{\rm lens}(z_i) = 0.055z_i$, as estimated by \citet{Jonsson2010} (for details on the other terms see section 3.2 of \citet{Scolnic2018}). $\mathbf{C}^{sys}$ is the covariance matrix induced by the training of the SALT2 model on a sample light curve set. 
\par 
Fitting was done using Polychord \citep{Handley2015}, where we fix the SN Ia fiducial absolute magnitude $M = -19.43$ which is equivalent to setting $h = 0.674$. We find $\Omega_{\rm m} = 0.298 \pm 0.022$, consistent with \citet{Scolnic2018} (although our sample is a little smaller) and we use the central value to compute the residuals. The standard deviation of the residuals was $\sigma_{\rm res} \simeq 0.14$ which combines intrinsic, lensing and all other sources of scatter.
\par
We compute the bin $k$ sample Pearson correlation coefficient $\rho_k$ between $\mu_{\rm res}$ and $\Delta m$ given in Equation (\ref{eq:normmag}) which is
\begin{equation}
\label{eq:pearsoncorrelation}
    \rho_k = \frac{\sum_{z_i \in (z_k, z_{k+1})} (\mu_{i, \rm res} - \langle \mu_{\rm res} \rangle) \Delta m_i}{\sqrt{\sum (\mu_{i, \rm res} - \langle \mu_{\rm res} \rangle)^2} \sqrt{\sum \Delta m_i^2} } \;,
\end{equation}
where the sum runs over all supernovae in bin $k$ and $\langle \Delta m_i \rangle = 0$ in each bin by construction. We will also calculate the correlation of the colour parameter $c$ and stretch $x_1$ with our lensing estimator $\Delta m$ to check for dust extinction and any lensing time delay.
\par 
As an additional cross-check, we also obtain the \textit{partial correlation coefficient} $r_{ij}$ between variable $i$ (that is, $\Delta m$, $\mu_{\rm res}$, $c$ or $x_1$) and variable $j$. The partial correlation is defined as the Pearson correlation coefficient between the residuals of the two variables of interest when the others have been fitted out by linear regression. Setting $\mathbf{\Omega} = (\rho_{ij})^{-1}$ it is given by 
\begin{equation}
    r_{ij} = - \frac{\Omega_{ij}}{\sqrt{\Omega_{ii} \Omega_{jj}}} \;.
\end{equation}

\subsection{Constraining halo parameters}
We are additionally interested in deriving Bayesian posteriors for the halo model parameters $\mathbf{\varpi}$. We use Bayes' theorem 
\begin{equation}
    P(\mathbf{\varpi} | \mathbf{x}, \mathcal{M}) = \frac{\mathcal{L_M} (\mathbf{\varpi}) \mathcal{\pi_M}(\mathbf{\varpi})}{\mathcal{Z_M}} \;,
\end{equation}
where $\mathbf{x} = \mathbf{\mu_{res}} - \mathbf{\Delta m}$ is our data vector, and $\mathcal{\pi_M} = P(\mathbf{\varpi} | \mathcal{M})$ is our prior belief in the parameters given the model $\mathcal{M}$. $\mathcal{Z_M} = P( \mathbf{x} | \mathcal{M})$ is the evidence given by integrating the likelihood $\mathcal{L_M} = P(\mathbf{x} | \mathbf{\varpi},\mathcal{M})$ over the prior, calculated as
\begin{equation}
    \mathcal{Z_M} = \int \mathcal{L_M} (\mathbf{\varpi}) \mathcal{\pi_M}(\mathbf{\varpi}) d \mathbf{\varpi} \; .
\end{equation}
We adopt the likelihood
\begin{equation}
\label{eq:likelihood}
    \ln \mathcal{L} = \mathbf{x}^{T} (\mathbf{C}')^{-1} \mathbf{x}
\end{equation}
where the adjusted covariance $\mathbf{C}'$ is derived from the Pantheon covariance by removing the stated lensing variance :
\begin{equation}
   \mathbf{C}' = \mathbf{C} - (0.055z_i)^{2}\;.
\end{equation}
Although in principle extra variance from uncertainties in our lensing calculation should be included, they are relatively small compared to the intrinsic SN Ia residual variation, and we can neglect them. Indeed, we find the distributional properties of $\mathbf{x}$ for our sample match this likelihood very well, with residual non-Gaussian properties small. 
\par 
We use uniform priors where the mass-to-light ratio $\Gamma \in (40,400)$ and the halo radial profile slope $\beta \in (0.5, 4.0)$, and when extending to a variable uniform concentration, the halo concentration $c \in (2,15)$. To derive our posteriors we use the nested sampling method implemented in Polychord \citep{Handley2015}. 
\par

\section{Results}
\label{sec:results}

\subsection{Description of the lensing signal}
In this subsection, we describe the features of the unnormalised magnification estimate given by Equation (\ref{eq:unnormmag}), with the M08 concentration model and maximum likelihood halo parameters $(\Gamma , \beta)$ (see \ref{tab:parameters}).
\par
\begin{figure}
    \centering
    \includegraphics[width=\columnwidth]{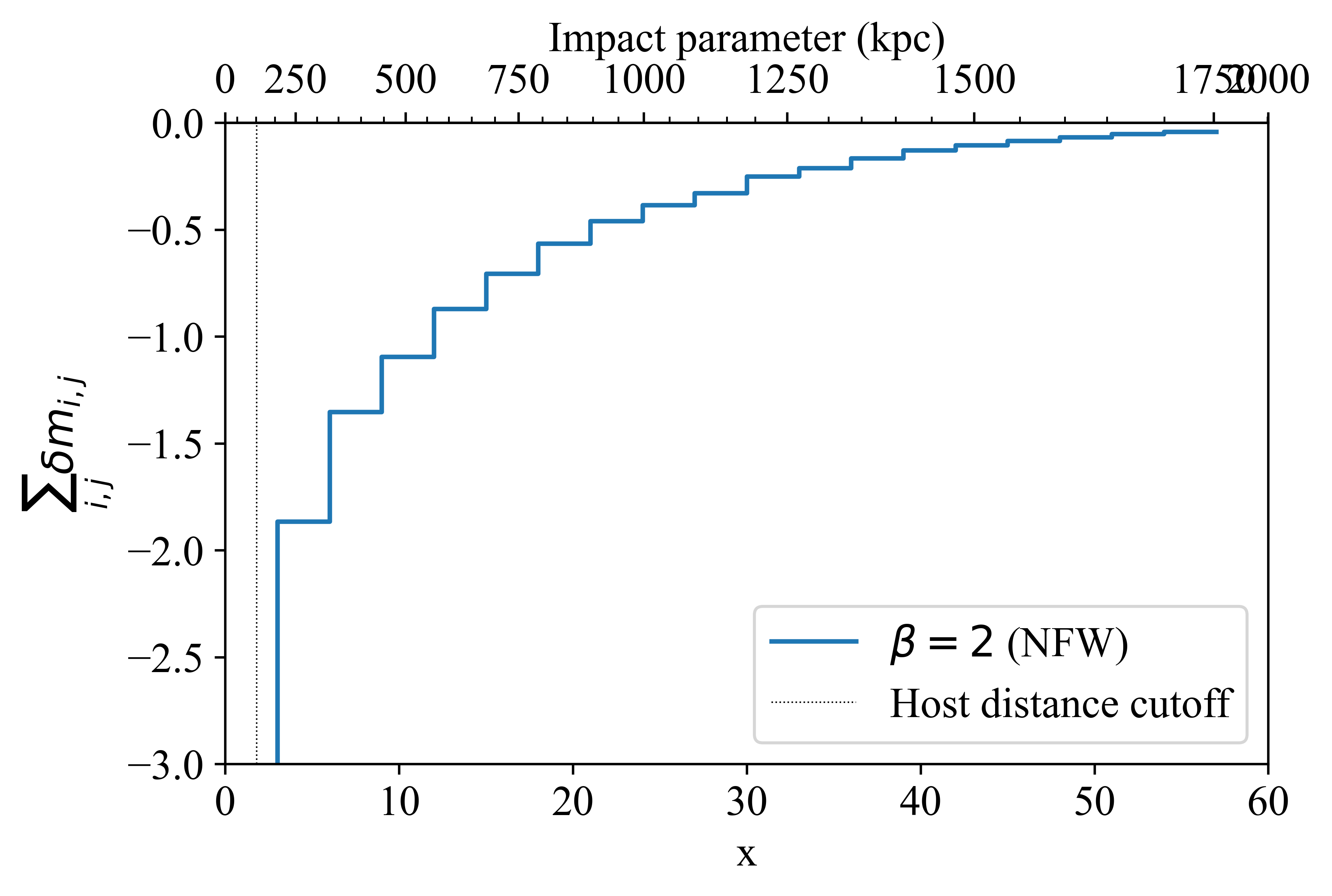}
    \caption{The aggregate unnormalised magnification of our sample, summed over all galaxies $j$ and SN Ia hosts indexed by $i$. The lower axis is the impact parameter normalised to units of the scale radius $x = b/r_{\rm s}$ where $r_{\rm s} = r_{200} / c$. The upper axis is the impact parameter $b$ averaged over the given $x$ bin. Our host identification criteria of the projected distance of the SN Ia to the nearest galaxy in scale radii $x<4$ is marked as a vertical black line at $r \sim 100$ kpc. Galaxies that are not SN Ia hosts will still contribute to our lensing calculation for $r<100$ kpc.}
    \label{fig:magxhistogram}
\end{figure}

The majority of lensing signal comes from galaxies whose impact parameters lie within scale radius $x \simeq 5 - 30$. For a typical Milky Way sized galaxy, this would be an impact parameter of $b \simeq 0.15 - 1.0 $Mpc. We illustrate this in Figure \ref{fig:magxhistogram}. Comparing this to our profiles from Figures \ref{fig:halobetaprofile} and \ref{fig:halocprofile}, we can see that we should be able to obtain moderate constraints on the slope parameter $\beta$, but that $c$ is unlikely to be constrained very well by our sample. 

\par
\begin{figure}
    \centering
    \includegraphics[width=\columnwidth]{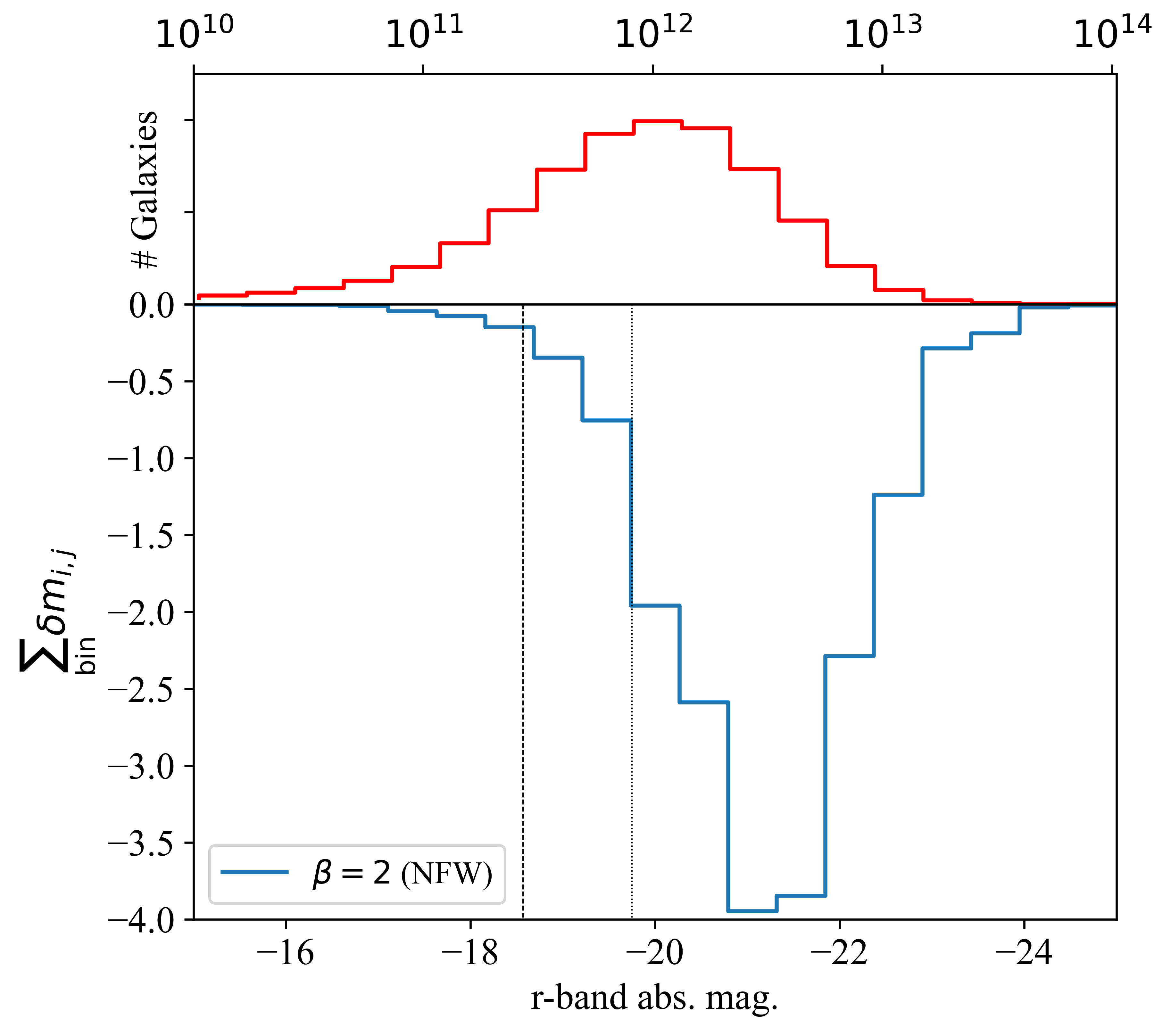}
    \caption{The upper panel shows the counts of galaxies in our sample binned by absolute magnitude $M_r$, with $M_{200}$ shown on the upper x-axis in units of $M_{\odot}$. The lower panel shows our total lensing signal summed over galaxies and binned by absolute magnitude of the galaxy lens. The majority of our signal is within the SDSS limiting magnitude $m_r = 22$, marked for redshift $z = 0.25, 0.4$ as the vertical black dashed and dotted lines.}
    \label{fig:magabsmag}
\end{figure}

In terms of which galaxies contribute to lensing, we show in Figure \ref{fig:magabsmag} the aggregate estimate bucketed by galaxy absolute magnitude. Although the numbers of galaxies peak at $M \sim -20$, the lensing signal peaks at $M \sim -21.5$ which is equivalent to a Milky Way-type galaxy. In the plot, we have marked the absolute magnitude of an $m_r = 22$ galaxy located at $z = 0.25, 0.4$. 

\par
\begin{figure}
    \centering
    \includegraphics[width=1.0\columnwidth]{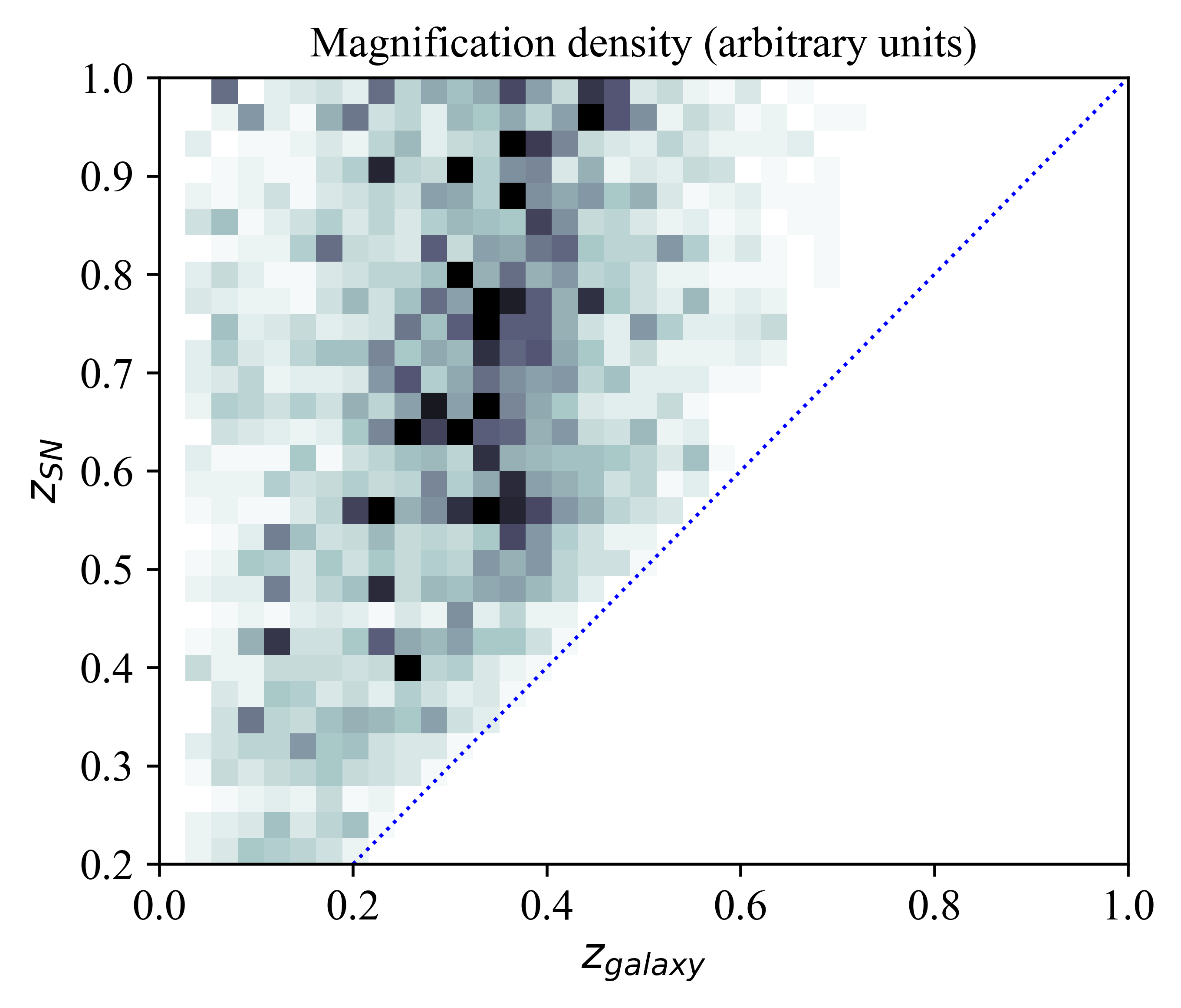}
    \caption{An illustration of the density of our lensing signal per SN Ia as a function of source redshift $z_{SN}$ and lens redshift $z_{\rm galaxy}$. Units are arbitrary, and are not shown. The lensing density peaks as expected; that is, roughly midway between source and observer. The effect of the galaxy magnitude limit is to reduce the density in the upper right hand corner of the plot.}
    \label{fig:magbysnredshift}
\end{figure}

In Figure \ref{fig:magbysnredshift}, we show an illustration by redshift where the lensing estimate peaks. As expected, it is generally midway in redshift between the SN Ia and $z=0$. For a SN Ia at $z \sim 0.5$ and a typical lensing galaxy at redshift $z \sim 0.25$, the magnitude limit $m_r = 22$ corresponds to $M_r = -18.5$, equivalent to the Large Magellanic Cloud. The mass of the LMC is $\sim 1/100$ of a typical galaxy, and so galaxies below the survey limit at such intermediate redshifts will contribute a relatively small amount to the overall lensing signal, even taking into account their larger number density. However, for SN Ia at  redshift $z \sim 1.0$ the typical lensing galaxy is $z \sim 0.4$ and so $M_r < -20.3$. We can now expect to be missing some fraction of the true lensing amount. We indeed see this in the top right of Figure \ref{fig:magbysnredshift} as the reduced density of the lensing signal close to the diagonal line, compared to low and intermediate redshifts. 

\subsection{Correlation of lensing estimate with SN Ia observables}

For our best fit model, we find a correlation between our lensing estimate $\Delta m$ and Hubble diagram residual $\mu_{\rm res}$ of $\rho = 0.166 \pm 0.045$(stat) for SN Ia with $0.2 < z < 1.0$. This is a significance of $3.7\sigma$ before allowance for systematics. 
\par 
The errors quoted have been derived from 10,000 bootstrap resamples of data. It is clear we should exclude low-z SN Ia, as we do not expect them to be measureably lensed. We also exclude $z>1$ SN Ia for several reasons. Firstly, we expect them to be lensed to a significant degree by galaxies below the magnitude limit of the SDSS survey. Secondly, they are drawn from surveys conducted by the Hubble Space Telescope (see for example \citet{Riess2007}), so the targetting and detection efficiency may differ considerably from ground-based surveys. In any case, there are not enough numbers of them for this exclusion to affect our result.  The result are not greatly changed for other concentration models; for the D08 model it is $\rho = 0.166 \pm 0.046$(stat), for MC11 it is $\rho = 0.166 \pm 0.046$(stat) and for a fixed $c=6$ it is $\rho = 0.151 \pm 0.048$(stat). 
\par 
Our other parameter choices were specified in Table \ref{tab:parameters}. As expected, the correlation drops if we do not exclude SN Ia with poorly identified hosts: for no exclusions, $\rho \sim 0.11$. We discuss the dependence on other analysis parameter choices in the next subsection.
\par 
For the NFW profile we find $\rho = 0.159 \pm 0.046$(stat), and for the SIS halo profile $\rho = 0.149 \pm 0.046$(stat). The dependence of the SIS correlation on aperture radius is greater than for the NFW model, and as expected the correlation declines with wider aperture as remote field galaxies dilute the lensing signal. Anticipating the results from the full Bayesian analysis described in the next section, profiles close to NFW are likely to be preferred to those close to SIS. The result for the $\beta = 3$ Hernquist profile is $\rho = 0.097 \pm 0.046$(stat).

\begin{figure}
    \centering
    \includegraphics[width=\columnwidth]{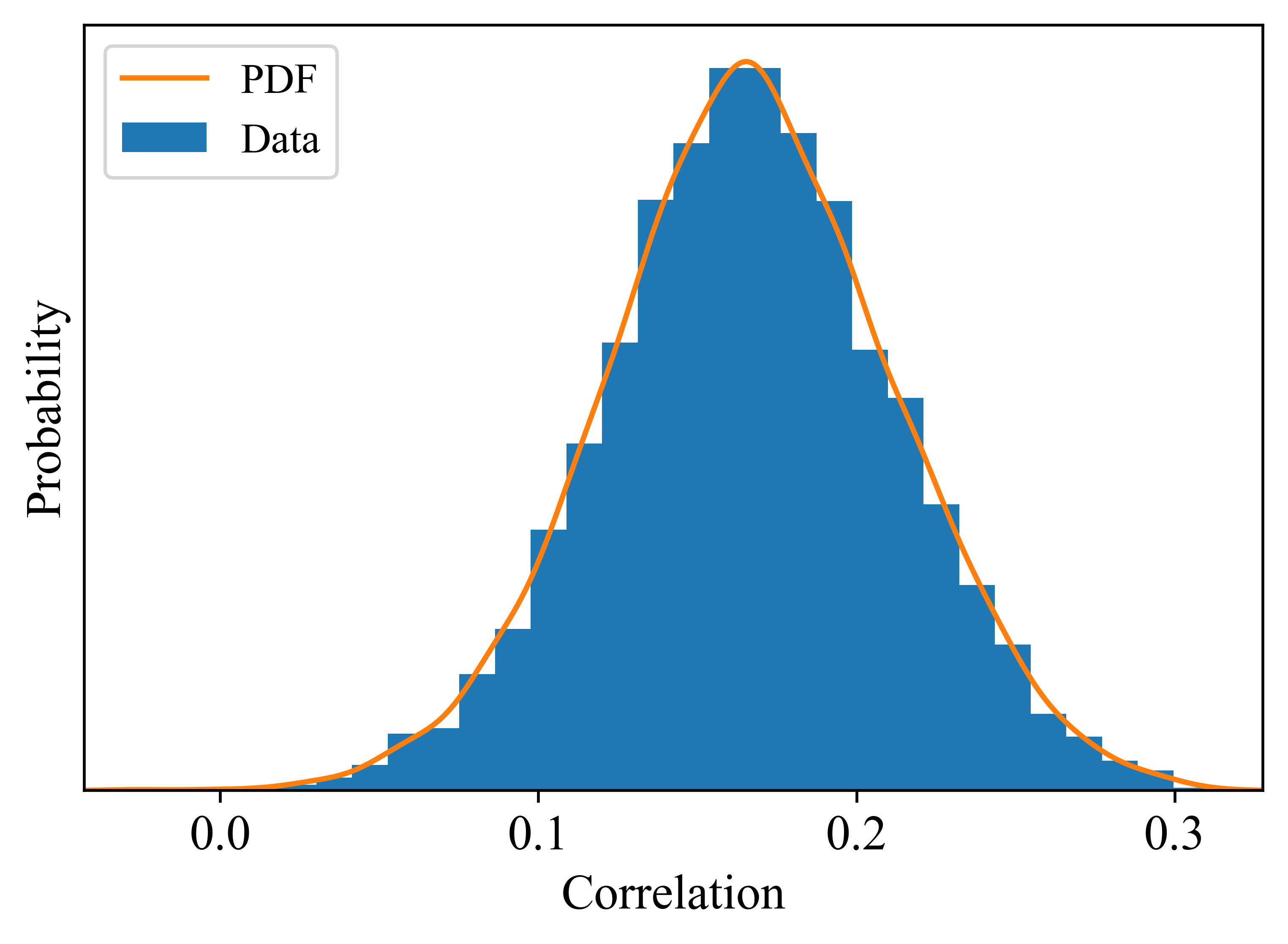}
    \caption{The bootstrap resampling distribution of correlation for our $0.2< z< 1.0$ sample. The statistical significance of lensing signal detection obtained is $\bar{\rho} / \sigma_{\rho}  = 3.6$. }
    \label{fig:corrbootstap}
\end{figure}

\par
\begin{figure}
    \centering
    \includegraphics[width=\columnwidth]{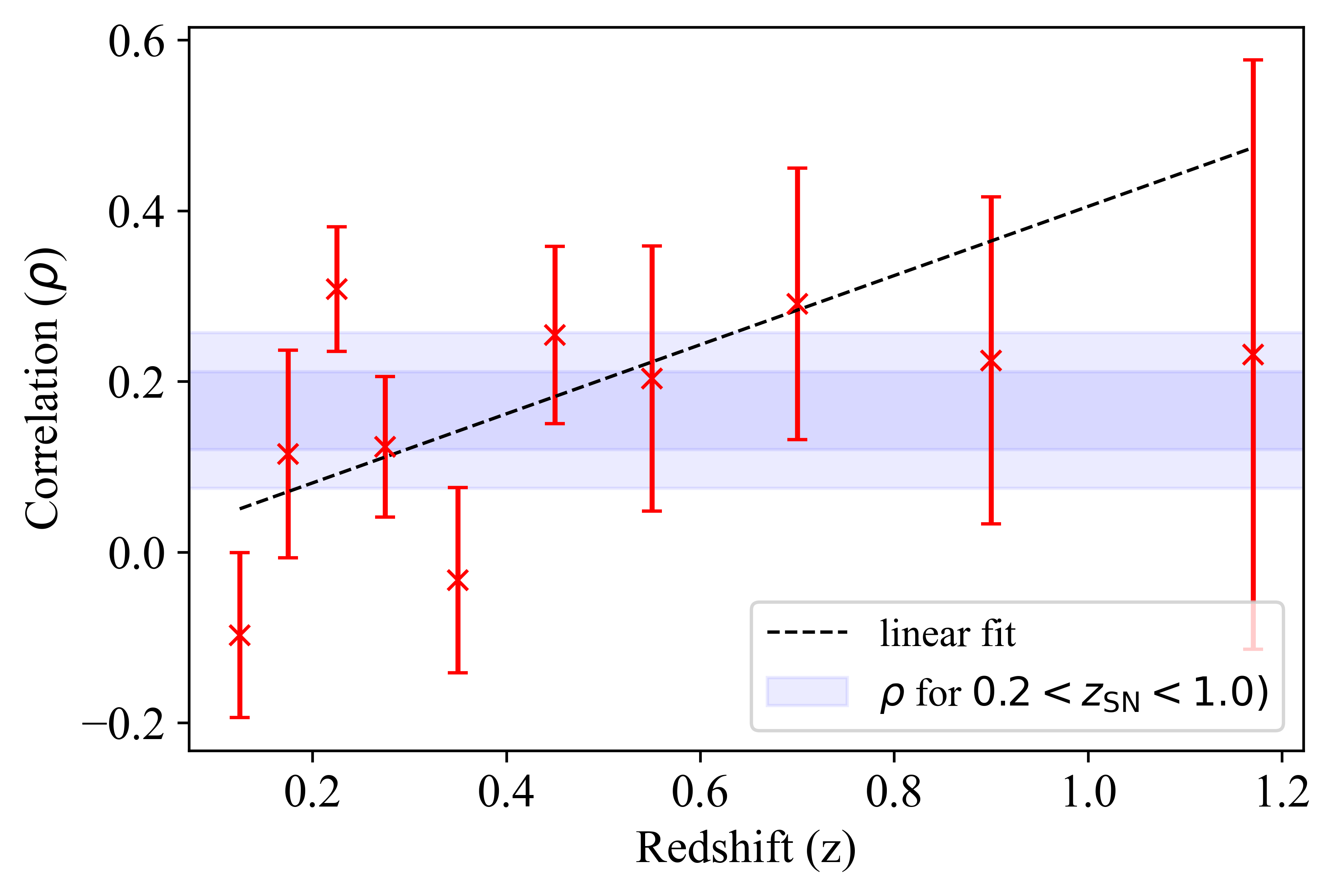}
    \caption{The correlation $\rho$ between the Hubble diagram residuals and weak lensing convergence estimate of our SN Ia sample, shown for individual redshift bins. Errors are computed by bootstrap resampling. The horizontal axis shows the average redshift in each bin. Our main result of $\rho = 0.166 \pm 0.046$ for the sample between $0.2 < z< 1.0$ is shown as the shaded purple bars at $1\sigma$ and $2\sigma$ confidence. As expected for a signal due to lensing, we see a generally increasing trend with distance, and a linear fit to the correlation is marked. The trend towards larger error bars with increasing redshift is due to the smaller numbers of SN Ia in distant bins. Additionally, for the bins $z>0.8$ the reduced significance is also due to the small angular field of the high-z HST surveys : there is not enough variation in the density of foreground galaxies across the field to show a correlation (this is discussed further in Section 5.5).}
    \label{fig:corrbucket}
\end{figure}

\begin{figure*}
    \centering
    \includegraphics[width=17cm]{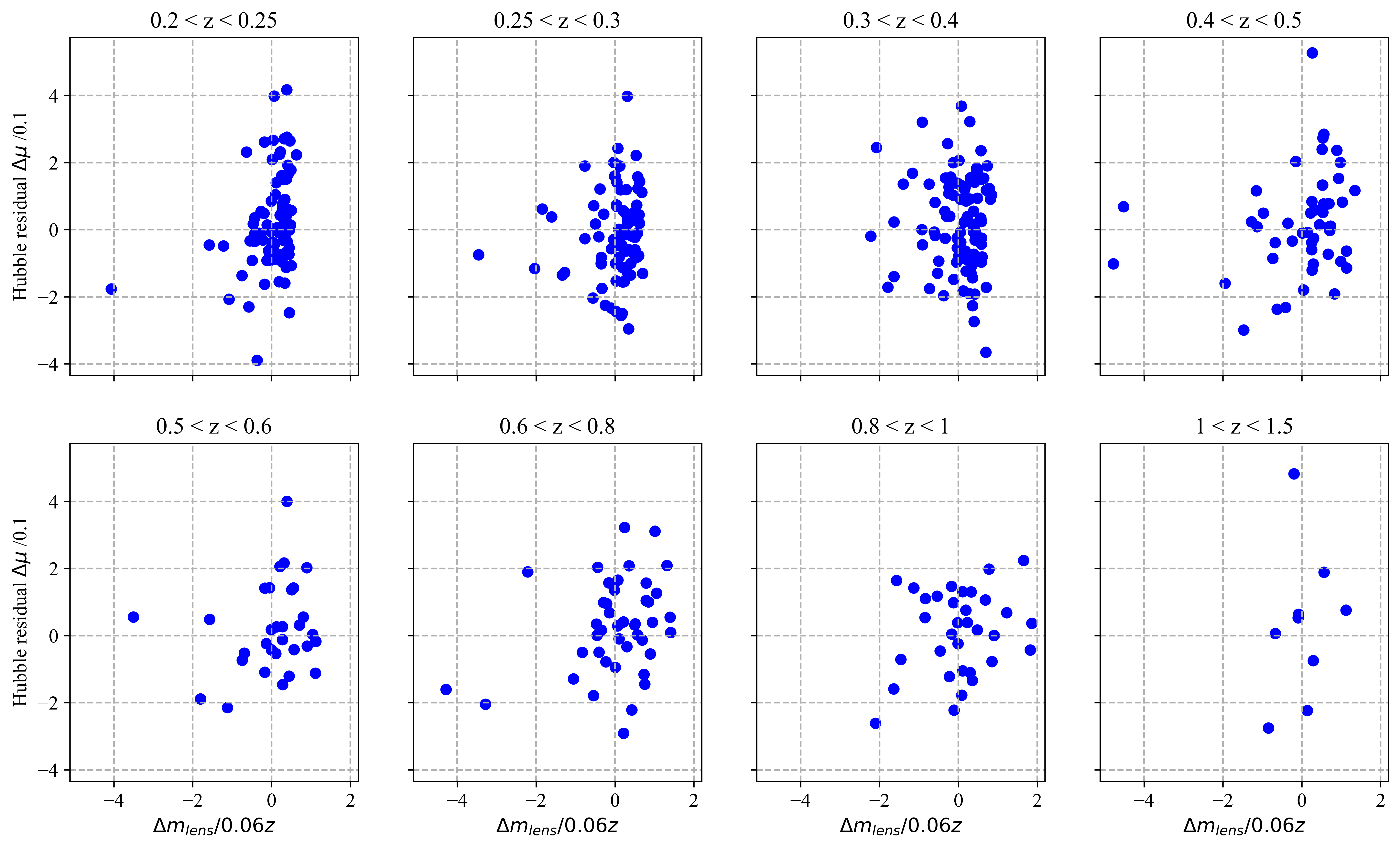}
    \caption{Scatter plots of Hubble diagram residuals $\mu_{\rm res} = \mu - \mu_{\rm model}$ of SN Ia (y-axis) and the lensing estimate $\Delta m$ (x-axis). We have normalised the scales by dividing by the expected dispersion $\sigma_{\rm lens}  = 0.06z$ and intrinsic dispersion $\sigma_{\rm int} = 0.1$. For low redshift bins, we see the intrinsic dispersion of magnitudes, which has low skew and $\sigma_{\rm int} \sim 0.1$. For higher redshift bins, the correlation is apparent as the clustering of points in the top right quadrant (the majority of lines of sight are through underdense regions) and a small number of magnified supernovae in the bottom left. A small number of  outliers are present; in particular, in the $0.6<z<0.8$ bin the point in the top left quadrant is SN 05D3hh (see text for comment).}
    \label{fig:magscatter}
\end{figure*}

 
In Figure \ref{fig:corrbucket} we show the correlation per redshift bin; as we would expect, we see a generally increasing trend with redshift, within the limitations of Poisson noise given $\sim 50$ SN Ia per bin. We show scatter plots of our residuals in Figure \ref{fig:magscatter}. As previously argued, the majority of our SN Ia are de-magnified and a smaller number of SN Ia are magnified. The intrinsic scatter dominates for low redshifts, but for larger redshifts the correlation is visible in the grouping of dots towards the bottom left and top right quadrants. Our correlation is lowered by a few outliers, notably the SNLS supernovae 05D3hh, 04D3nr, 05D3km, 05D3mh and 04D3gx which are points well inside the upper left quadrants in the bins $0.6<z<1.0$. These are SN Ia that have foreground galaxies close to the line of sight but are dimmer than the Hubble diagram fit. Four out of five of these were originally classified by the survey as \say{probable} (rather than certain, see \citet{Conley2011}) SN Ia due to some ambiguity in their spectral classification. The proportion of probable SN Ia in the survey is $\sim 20\%$, so it is possible they represent contamination of the sample by non-SN Ia. Nevertheless, they have passed Pantheon quality cuts for their light curve fitting, and there is no objective reason to exclude them. Without the additional criteria of compatibility of redshift between the host galaxy and SN Ia, the correlation is $\rho = 0.177 \pm 0.046$. Hence the effect of the outliers is to modestly reduce the statistical significance from $3.9\sigma$ to $3.7\sigma$.

\par

\par 
The correlation of our lensing estimate with colour is not significant at $\rho_{\rm c} = 0.024 \pm 0.05$. This result means we have no evidence for dust extinction in galactic halos. This is consistent with \citet{Smith2014}, who also found no correlation between colour and their lensing estimate, but does not contradict the results of \citet{Menard2010a}. The reason for this is that we find the average colour variation for our sample, as computed by equation (\ref{eq:deltaci}), to be $\Delta c \sim 0.002$. Whereas for magnification, our lensing signal is between $10\%$ to $50\%$ of the intrinsic variation, for colour our signal is just $2\%$ of the intrinsic variation of $\sigma_c \sim 0.1$. We would therefore only expect dust extinction to be detected only if we had a sample exceeding $10,000$ SN Ia. 
\par 
The correlation between lensing and stretch is not significant at $\rho_{x_1} = 0.047 \pm 0.047$ (in the literature, \citet{Smith2014} have reported a $2.2 \sigma$ significance). As a further test, we confirm the partial correlation coefficients are consistent with our results. 
\par
We summarise our results on correlation in Tables \ref{tab:pearsoncorr} and \ref{tab:partialcorr}.
\begin{table}
    \centering
    \begin{tabular}{c c c c} \hline\hline
     & $c$ & $x_1$ & $\Delta m$ \\ \hline
    $\mu_{\rm res}$ & -0.020(0.3) & -0.027(0.5) & \textbf{0.166(3.6)} \\
    $c$ & ... & 0.025(0.5) & 0.024(0.5) \\
    $x_1$ & ... & ... & 0.047(1.0) \\ \hline
    \end{tabular}
    \caption{Pearson correlation coefficient between the lensing signal $\Delta m$, SN Ia Hubble diagram residual $\mu_{\rm res}$, colour $c$ and stretch $x_1$. The significance of each correlation (including systematics) is also given in brackets. Our main result is highlighted in bold.}
    \label{tab:pearsoncorr}
\end{table}

\begin{table}
    \centering
    \begin{tabular}{c c c c} \hline\hline
     & $c$ & $x_1$ & $\Delta m$ \\ \hline
    $\mu_{\rm res}$ & -0.023 & -0.035 & 0.168 \\
    $c$ & ... & 0.024 & 0.026 \\
    $x_1$ & ... & ... & 0.051 \\ \hline
    \end{tabular}
    \caption{Partial correlation coefficient between the lensing signal $\Delta m$, SN Ia Hubble diagram residual $\mu_{\rm res}$, colour $c$ and stretch $x_1$. The results are largely unchanged compared to the Pearson correlation.}
    \label{tab:partialcorr}
\end{table}

\subsubsection{Correlation systematics}
We test the robustness of our results to our parameter choices; our correlation is unlikely to be over-estimated by a \say{bad} parameter choice, but we seek to estimate a systematic error to complement our statistical errors. 
\par 
We perform the following tests on photometric selection criteria. We vary the passband used to calculate the absolute magnitude (and hence mass) of the lensing galaxies. We test the effect of the varying the magnitude limit for galaxies above and below the formal survey limit at $m_r = 22$. As we would expect, if we adopt a lower limit we exclude some galaxies that would contribute to the lensing signal, and the correlation drops. We additionally test the effect of varying our criteria for accepting heavily masked fields. If this selection parameter is too low, we drop too many SN Ia and the correlation drops. Alternatively, adopting too many masked fields adds noise to the signal and again the correlation drops. We find our chosen cut of $50\%$ works well. 
\par
Photo-z errors will affect all physical distances used in our calculation, including angular diameter distances, the impact parameter $b = D_{\rm d}(z) \theta$ and the critical surface density $\Sigma_c(z)$. As the convergence $\kappa \;\propto\; 1/b^{\beta}$, it is immediately clear that a bias might be introduced into our lensing calculation, even if the underlying photometric redshifts are themselves unbiased. In particular, the steeper the halo profile, the larger the potential bias. Further, it is likely photo-z errors will be correlated to some degree (given the size of the training set relative to the survey size), but the degree of such covariance is difficult to estimate.
\par
We address photo-z errors by multiplying each $z$ by a random lognormal error of width $\sigma_z /(1+z)$, and rerunning our analysis.  We also test the relatively extreme scenario of always multiplying or dividing by the relative error - this is intended to determine the effect of fully correlated errors in redshifts. We additionally test the exclusion criteria for poor quality redshifts by varying the maximum and minimum photo-z error class we select.
\par
Adopting the average of the change in our correlation across our choices as an estimate of potential systematics, we find $\sigma_{\rho} = 0.011$(sys). Adding this in quadrature to our statistical error gives our main result $\rho =0.166 \pm 0.046$ (stat+sys). This is a detection significance of $3.6\sigma$.

\subsection{Halo parameters}
We find for the M08 model mean values of $\beta = 1.8 \pm 0.3$ and $\Gamma = 197 ^{+64}_{-80} \, h \, M_{\odot}/L_{r, \odot}$ where marginalised $65\%$ confidence intervals are indicated\footnote{We have re-introduced $h= 0.674$ to normalise our result here.}. The posteriors on our model parameters imply an additional error on $\rho$ of $\sigma_\rho = 0.007$(post). For comparison, the maximum likelihood values are $\beta = 1.7$ and $\Gamma = 196 \, h \, M_{\odot}/L_{r, \odot}$. Fixing $\beta =2$, the NFW profile gives $\Gamma = 224 ^{+58}_{-79}\, h \, M_{\odot}/L_{r, \odot}$.
\par
We find our results are again largely unaffected by the choice of concentration model; the D08 model prefers a slightly higher $\beta$ as its average concentration is lower than M08, and conversely the MC11 model prefers a lower $\beta$ for the same reason. 
\par
\begin{figure}
    \centering
    \includegraphics[width=\columnwidth]{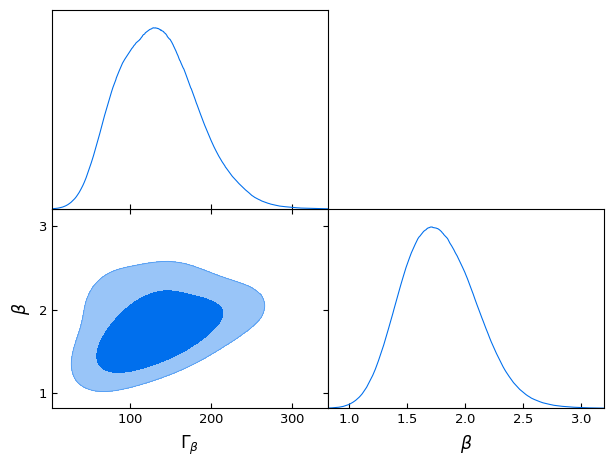}
    \caption{The derived posteriors for our power-law halo profile slope $\beta$, and mass-to-light ratio $\Gamma_{\beta}$. $\beta = 2$ corresponds to the NFW profile. We also show the correlation posterior implied by the range of halo profiles; our core result of lensing detection is largely insensitive to the details of the profile. As explained earlier, there is some degeneracy between concentration determined by either $\beta$ or $c$, and $\Gamma_{\beta}$ : a more concentrated halo profile requires a higher mass to produce the same amount of magnification.}
    \label{fig:triangleplot}
\end{figure}

We find for the power law model that $\beta > 1.2$ at $95\%$ confidence regardless of the concentration model used, and thus $\beta = 1$ (the modified SIS profile) is disfavoured to high confidence. We are also able to rule out the Hernquist profile at $>95\%$ confidence. The posteriors are illustrated in Figure \ref{fig:triangleplot}. As discussed in Section 2 in the context of concentration, we see some degeneracy between $\beta$ and $\Gamma$, whereby a higher $\beta$ favours more massive haloes to produce equivalent lensing power. 

\par
We test the effect of allowing a (uniform) $c$ to vary globally across our galaxy sample. For the NFW model, we find $c = 4.3 ^{+1.0}_{-2.0}$. This value is consistent with our concentration models above. We also test running a loosely constrained model where all of $(\Gamma, \beta, c)$ are allowed to vary. In this case, we find marginal values of $\Gamma = 185 ^{+55}_{-80}\, h \, M_{\odot}/L_{r, \odot}$, $\beta = 1.8 ^{+0.2}_{-0.6}$ and $c = 7.5 ^{+6.3}_{-5.3}$. These values are consistent with our main result above, although of course the confidence intervals are wider.
\par
As a check to derive evidences for model comparison, we also run the SIS profile and derive a posterior for $\Gamma = 70 ^{+4}_{-11}\, h \, M_{\odot}/L_{r, \odot}$.
\par
The results for our runs are presented in Table \ref{tab:posteriors}. Comparing the Bayesian evidence, we find little preference between our different concentration models (first four lines). The SIS and NFW models are somewhat disfavoured compared to allowing a variable slope; to a certain extent this is due to the presence of the small number of outliers (see Figure \ref{fig:magscatter}), otherwise we would find $\beta$ aligned with the NFW model. 

\begin{table*}

\resizebox{\textwidth}{!}{
\begin{tabular}{l|l||c c c|c c c|c c c||c c}
    \bfseries Profile & \bfseries Conc. & \bfseries $\Gamma$ & \bfseries 65\%  & \bfseries 95\%  & \bfseries $\beta$ & \bfseries 65\%  & \bfseries 95\%  & \bfseries $c$ & \bfseries 65\%  & \bfseries 95\% & \bfseries $\log{\mathcal{Z}}$ & \bfseries 65\% 
    \begin{filecontents*}{Polychord_results4.csv}
python,output,aperture,live,masslight (starter),nfwc,mingal,distmax,mmf,useband,pzmin,pzmax,profile,cmodel,zmin,zmax,gammaprior,betaprior,cprior,gammamean,gammaone,gammatwo,betamean,betaone,betatwo,cmean,cone,ctwo,logz,logzerr,gammameanh,minus,plus
lensing_beta_for_gamma_beta,lens_50r8_Mandelbaum_Beta,8,50,150,6,0,4,0.5,r,-3,7,Power law,M08,0.2,1,40/400,0.5/4,-,133,79/176,46/229,1.8,1.4/2.1,1.2/2.4,-,-,-,688.3,0.2,197,-80,64
lensing_beta_for_gamma_beta,lens_50r8_Duffy_Beta,8,50,150,6,0,4,0.5,r,-3,7,Power law,D08,0.2,1,40/400,0.5/4,-,140,84/185,51/238,1.9,1.6/2.2,1.2/2.6,-,-,-,688.6,0.2,208,-83,66
lensing_beta_for_gamma_beta,lens_50r8_Cuartas_Beta,8,50,150,6,0,4,0.5,r,-3,7,Power law,MC11,0.2,1,40/400,0.5/4,-,121,72/158,41/209,1.7,1.4/1.9,1.2/2.2,-,-,-,688.4,0.3,180,-73,54
lensing_beta_for_gamma_beta,lens_50r8_Fixedc6_Beta,8,50,150,6,0,4,0.5,r,-3,7,Power law,Fixed,0.2,1,40/400,0.5/4,6,164,50/211,36/355,2.1,1.6/2.4,1.4/3.1,6,-,-,688.9,0.2,243,-169,70
lensing_nfw_for_gamma_c,lens_50r8_Fixed_Nfw_c,8,50,150,6,0,4,0.5,r,-3,7,NFW,Fixed,0.2,1,40/400,-,2/15,134,85/177,49/228,2,-,-,4.3,2.3/5.2,1.6/8.4,688.3,0.2,199,-73,64
lensing_SIS_for_gamma,lens_50r8_SIS,8,50,150,6,0,4,0.5,r,-3,7,SIS,-,0.2,1,40/400,-,-,47,40/50,38/65,-,-,-,-,-,-,689.5,0.3,70,-11,4
lensing_beta_for_gamma,lens_50r8_Mandelbaum_Beta2,8,50,150,6,0,4,0.5,r,-3,7,NFW,M08,0.2,1,40/400,-,-,151,98/190,64/249,2,-,-,-,-,-,689.3,0.1,224,-79,58
lensing_beta_for_beta_c,lens_50r8_Fixed_Beta_c,8,50,150,6,0,4,0.5,r,-3,7,Power law,Fixed,0.2,1,40/400,0.5/4,2/15,125,71/162,38/227,1.8,1.3/2.0,1.1/2.7,7.5,2.2/13.8,1.8/14.6,688.6,0.2,185,-80,55
\end{filecontents*}
\csvreader[head to column names]{Polychord_results4.csv}{}
    {\\\hline \profile & \cmodel & \gammamean & \gammaone & \gammatwo & \betamean & \betaone & \betatwo & \cmean & \cone & \ctwo & \logz & \logzerr } 
\end{tabular}
}

\caption{A summary of the marginalised mean values and confidence intervals for $\Gamma$, $\beta$ and $c$, for a range of models and fixed or free parameters. Conc. refers to the concentration model used. $\Gamma$ is in physical units of $M_{\odot}/L_{r, \odot}$, but is not normalised by $h$ here. $\log{\mathcal{Z}}$ is the log of the mean likelihood as output by Polychord. The first three rows correspond to different concentration models. The fourth row shows the result when $c=6$ fixed - as this lower than is preferred in the free fit shown in row five so, $\Gamma$ is pushed high as a result. The sixth to ninth rows show $\Gamma$ when the profile and concentration model is fixed. The last row shows a 'global' result where all parameters are allowed to vary, and is shown as a consistency check. Priors were $\Gamma \in (40,400)$ and $\beta \in (0.5, 4.0)$, and when extending to a variable fixed concentration, $c \in (2,15)$.}
\label{tab:posteriors}
\end{table*}

\subsection{Malmquist bias}

Both our galaxy and supernovae surveys are magnitude-limited, and and we estimate its effect on our results here.
\par 
The effect of the magnitude limitation $m_{\rm gal}$ of the galaxy survey is straightforward to understand. A SN Ia will be lensed by all galaxies along the line of sight, whether seen or unseen. The redshift of a given SN Ia defines a volume limit applicable to foreground galaxies. Then, for SN Ia surveys paired with galaxy surveys on similar platforms, we will observe all contributing galaxies brighter than the SN Ia (and generally better than that, if the galaxy survey is deeper). Hence we will be largely unaffected by the galaxy survey limit. However, the SNLS and HST surveys are deeper than SDSS, and we will under-estimate the number of galaxies contributing to lensing. Hence the signal-to-noise and correlation will be lower, and the mass of low-z galaxies (that is, the surface density attributed to them) will be over-estimated.  
\par 
The effect of the magnitude limitation $m_{\rm l}$ of our supernova survey is more complex. For a magnitude limited sample, a small number of sources that would have been too faint to be seen in a homogeneous universe will magnify in to the observed sample (and will be identified as over-luminous SN Ia by their redshifts). Conversely, along under-dense lines of sight, SN Ia that are close to the magnitude limit will drop out of our sample. Taking the magnified SN Ia in isolation, flux conservation no longer holds : they represent over-dense lines of sight not representative of the homogeneous average. As their brightnesses sample the high-magnification tail of the lensing distribution, they will also show a greater dispersion. Hence, we can expect to see a spike in the dispersion of magnitudes in the redshift bucket containing $z(m_{\rm l})$ of the SN Ia sample. 
\par 
A practical difficulty for SN Ia is that $m_{\rm l}$ is not well-defined. The Pantheon compilation merges surveys with different limiting magnitudes. Further, only a small proportion of candidate supernovae (sometimes as little as 1 in 100) are targeted for spectroscopic followup, and the transient nature of the source means the decision-making process may be influenced by many factors such as instrument availability, local seeing conditions, SN Ia environment and so on, as well as magnitude. Hence there is no $m_{\rm l}$, but instead a \textit{detection efficiency} $f \in (0,1)$ is defined as the ratio of SN Ia that will be in the spectroscopically selected sample, compared to one with no selection. $f$ is determined by a model of the targeting algorithm and intrinsic scatter, and is thought to be well-characterised by the observed magnitude $m$ such that $f = f(m)$ (see for example the discussion in Section 3.3 and Figure 6 of \citet{Scolnic2018}). However, if a preference for a \say{clean} line of sight influences the selection, a sample biased towards under-dense lines of sight will result. 
\par
Taking the limits as where the survey is $\sim 50\%$ complete, we set $m_{\rm gal} = 22.5$, $m_{\rm SDSS} = 22.5$, $m_{\rm PS1} = 23.0$ and $m_{\rm SNLS} = 24.3$ (we will not use HST data for reasons discussed in the next subsection). Interestingly, in Pantheon the average stretch and colour parameters $\bar{c}, \bar{x_1}$ drift with higher redshift towards brighter SN Ia, and are most different from their mean of zero for the sharply truncated SNLS survey (see Figure 10 of \citet{Scolnic2018}). If this is due to selection bias, it is probable lensing will have a similar effect (however, the drift might instead be attributed to population evolution at higher redshift \citep{Nicolas2021}).
\par 
We tested the effect of these magnitude limits using the pencil-beam lightcones of \citet{Henriques2012} which are magnitude complete to $z \sim 1$, derived from the Millenium simulation \citep{Guo2010, Lemson2006, Springel2005}, and sourced from the German Astrophysical Virtual Observatory\footnote{\url{http://gavo.mpa-garching.mpg.de/Millennium/}}. As a cross-check we also used the broader MICE-Grand Challenge Galaxy and Halo Light-cone Catalog (Micecatv2.0) which is complete to $i<24$ \citep{Fosalba2015, Crocce2015, Fosalba2015a,  Carretero2015, Hoffmann2015}, sourced from Cosmohub\footnote{\url{https://cosmohub.pic.es/home}} \citep{Tallada2020, Carretero2018}.
\par 
Using random lines of sight and a fiducal $\Gamma = 150$, we examined the effect of imposing galaxy magnitude limits. The effect was to raise $\Gamma$ by $15\%$ (SNLS), $5\%$ (PS1) and negligible change for the SDSS SN Ia survey. As a further test of the galaxy magnitude limit, we re-ran Polychord for Pantheon and SDSS real data restricted to redshift $0 < z < z_{\rm max}$ for $z_{\rm max}$ between $0.5$ and $1.0$. This transitions our data sample towards (but not fully) volume-limited rather than magnitude-limited. We found a a modest drift upwards in the maximum-likelihood $\Gamma$ for lower $z_{\rm max}$, by about $8\%$ for $z_{\rm max}=0.5$. This is consistent with the results from simulations above.
\par
We next tested the effect of imposing the SN Ia survey limit. The dispersion of lensing was increased in the redshift bucket including the survey limit. As a result of the bias to over-dense lines of sight, $\Gamma$ was lowered by about $15\%$ for SDSS, $5\%$ for PS1 and unchanged for SNLS. 
\par 
Hence, the combined effect of the SN Ia and galaxy magnitude limits is somewhat offsetting in our data. By appropriately weighting the biases according to the number counts of each survey, we find Malmquist bias is expected to be $< 10\%$ on our derived parameters, well within the $1\sigma$ confidence intervals for the parameters we derive. We therefore do not adjust our fits. 

\subsection{Lensing dispersion}
\begin{figure}
    \centering
    \includegraphics[width=\columnwidth]{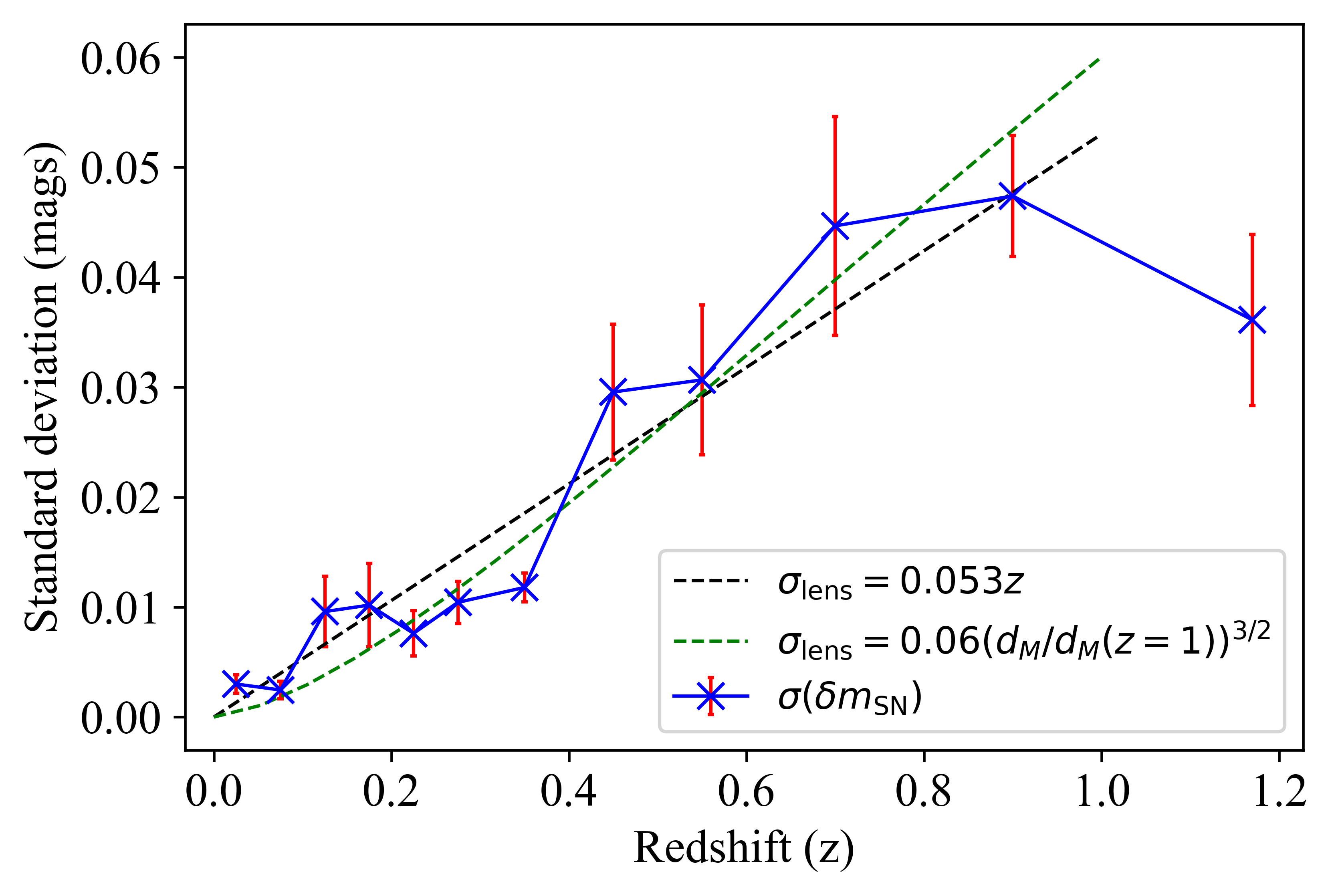}
    \caption{The dispersion of SN Ia magnitudes due to lensing. The error bars have been computed by bootstrap resampling. For $z < 0.8$, we estimate $\sigma_{\rm lens} = 0.053 z$. This does not extrapolate well, and we introduce the new fit $\sigma_{\rm lens} = 0.06(d_{\rm C}(z)/d_{\rm C}(z=1))^{3/2}$ which is better motivated. The outlier in the $1<z<1.5$ bucket is due to the limited numbers of SN Ia in this bucket being within a few arcminutes of each other.}
    \label{fig:lensingdispersion}
\end{figure}

We present the dispersion of lensing for the $\beta = 1.8$ model with M08 concentration and mean $\Gamma=133 \, M_{\odot}/L_{r, \odot}$ in Figure \ref{fig:lensingdispersion}. At redshifts $z>0.8$, the dispersion starts to drop below trend as expected due to magnitude limits. We also show the dispersion for the $1<z<1.5$ bucket, which seems anomalously low. We only have 9 SN Ia that pass our quality criteria in the bucket $1.0 < z < 1.5$, and all are from the HST surveys GOODS, CANDELS and CLASH. There is a straightforward explanation : these were pencil-beam surveys, and most of the SN Ia are within a few arcminutes of each other. There will then be little variation between lines of sight!
\par
The standard deviation is sensitive to the high magnification tail. \citet{Jonsson2010} argued that due to Poisson noise and the limited size of their sample (175 SN Ia), they were missing highly magnified $\Delta m < -0.25$ supernovae that would increase the dispersion. The authors replaced the dispersion from their actual sample $\sigma_{\rm lens} = 0.035z$, with $\sigma_{\rm lens} = 0.055z$ which was the dispersion from their best-fit model across a large number of randomly selected lines of sight. We are less likely to be affected by Poisson noise as we have higher numbers of SN Ia; Figure \ref{fig:magscatter} shows there are adequate numbers of SN Ia with $\Delta m < -0.25$ for $z<0.8$ but less so for the last two buckets. We therefore restrict our fit to $z<0.8$ where the bias is small. 
\par 
It is usual to fit for $\sigma_{\rm lens} = A z$, perhaps because the data appears visually linear. If we do so, we find $A = 0.053 \pm 0.015$. This is consistent with \citet{Jonsson2010}, and also with \citet{Bergstrom2000a} who found $\sigma_{\rm lens} = 0.04z$, but lower than \citet{Holz2005} who estimated $\sigma_{\rm lens} = 0.088z$, both from simulations.
\par
Obviously this fit has a limited range, and extrapolating it beyond $z>1$ is dubious. In Appendix B, we show that in the case of no clustering of galaxies\footnote{This is equivalent to the \say{stochastic} approaches of authors such as \citet{Holz2005,Jonsson2010, Marra2013}.} a better fit is provided by 
\begin{equation}
    \sigma_{\rm lens} = B (d_{\rm C}(z)/ d_{\rm C}(z=1))^{3/2}
\end{equation}
where we have normalised to the comoving distance at $z=1$, and $d_{\rm C}(z=1)= 3400 \; \mbox{Mpc}$ for our fiducial cosmology. We find $B = 0.06 \pm 0.017$ and the fit is shown on Figure \ref{fig:lensingdispersion}. This is likely to be a \textit{lower bound} on the true dispersion, due to both the magnitude limitation of the survey and the effect of clustering (which introduces extra covariance between sources and lensing galaxies). We recommend to use this in cosmological parameter estimation, as it is generally higher than previously assumed values. 
\par 
As a cross-check, we simulated 10,000 lines of sight from randomly selected galaxies in the SDSS footprint, and recomputed the lensing variance. We did not find any significant difference from the above, and therefore conclude our fit is unaffected by shot noise. 

\subsection{Comparison with shear studies}
Direct comparisons of the mass-to-light ratio we derive to the literature are complicated by the selection of which luminosity to compare to which mass, and also differing halo profiles and truncations. For example, the comparison may be between the luminosity of the single brightest galaxy in a cluster to the mass of the entire cluster. 
\par
\begin{figure}
    \centering
    \includegraphics[width=0.8\columnwidth]{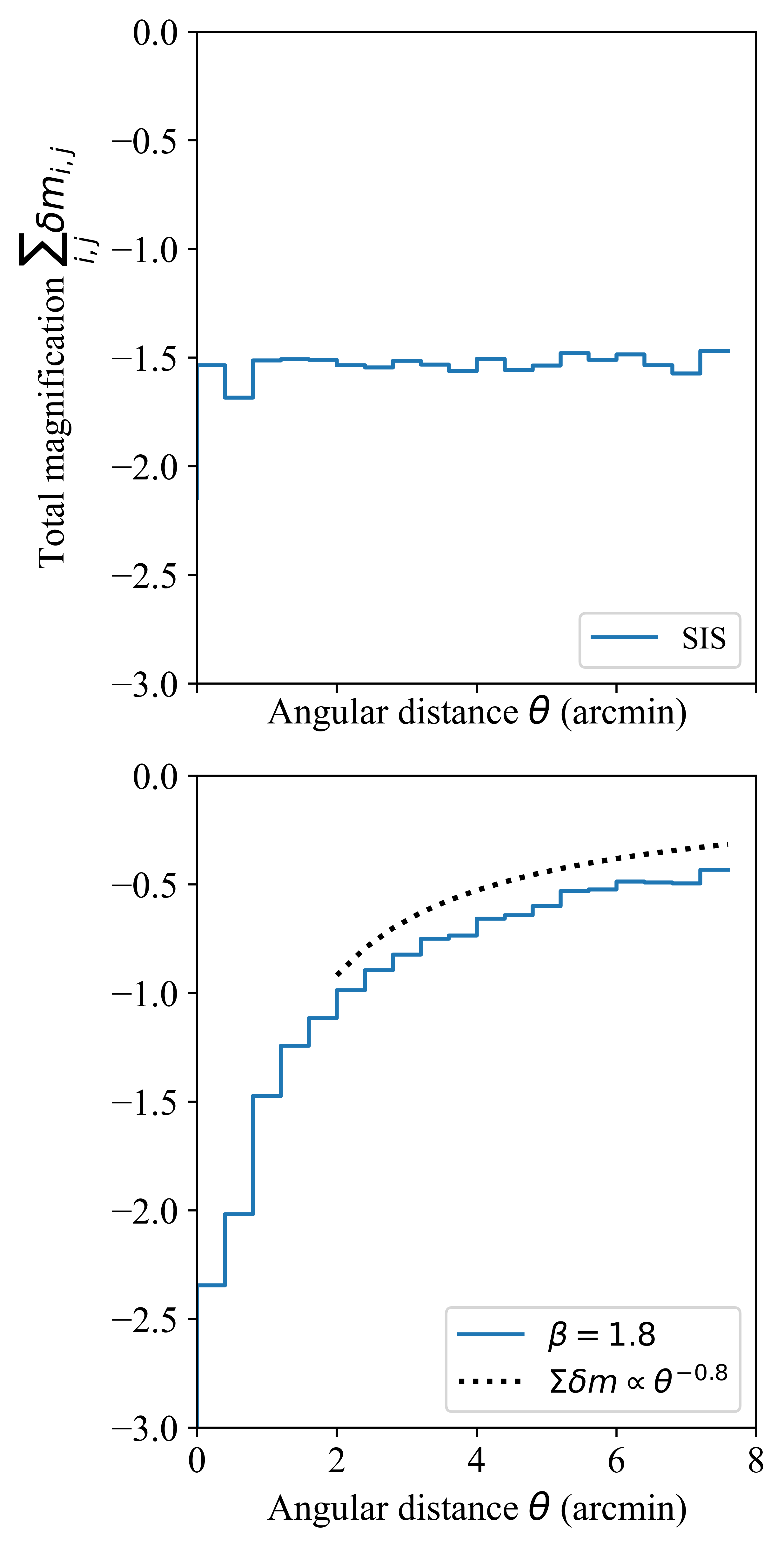}
    \caption{Our total lensing signal summed over galaxies and binned by angular distance $\theta$ from the SN Ia host. In the left plot, we show the SIS halo calculation. We obtain a flattish profile as the increase in the numbers of galaxies within each increasing size of annulus of sets the lower surface density with larger separation. In the right hand graph, we show the best-fit $\beta = 2.14$ profile. In this case, the lensing signal is well-fit by $\theta^{-1}$, which is consistent with shear studies.}
    \label{fig:magtheta}
\end{figure}

Regarding the radial dependence of the convergence, \citet{Menard2010a} studied the magnification of quasars by galaxies drawn from the SDSS survey, finding the projected $\Sigma (r) \; \propto \; r^{-0.8}$ from 10 kpc to 10 Mpc. This is consistent with other shear studies (see for example \citet{Sheldon2004}). We find that after stacking our lensing estimate into angular buckets, the power law model with $\beta = 1.8$ fits $\Sigma(\theta) \; \propto \;\theta^{-0.8}$. This is shown in Figure \ref{fig:magtheta}. This is because a SIS profile for a total surface density surrounding a fiducial galaxy is mimic-ed by the overlapping contributions of nearby NFW halos. For luminous red galaxies from the SDSS survey, \citet{Mandelbaum2006} found the NFW profile was preferred to the SIS at a confidence level of $96\%$, and found an average $c = 5.3 \pm 1.2$.
\par 
There is good evidence that mass-to-light ratios depend on galactic morphology and colour. The value we derive here should be seen as a population average, weighted by luminosity. In a shear study, \citet{VanUitert2011} compare the total luminosity within $r<r_{200}$ to the mass $M_{200}$ for an NFW profile with the D08 concentration model. For bright early-type galaxies with $L > 5 \times 10^{11} L_{\odot}$ they find $\Gamma \sim 260 \, h \, M_{\odot}/L_{r, \odot}$, but considerably lower values for late-types or lower luminosities. In a galaxy-galaxy lensing study, \citet{Brimioulle2013} find $\Gamma = 178^{+22}_{-19} \, h \, M_{\odot}/L_{r, \odot}$ at a reference luminosity of $L_{r} = 1.6 \times 10^{10} \, h^{-2} \, L_{r, \odot}$. 
\par
We may also compare our value to a \say{cosmic} mass-to-light ratio; that is to say the value approached on large scales. \citet{Bahcall2014} examine shear around SDSS clusters and derive a cosmic $\Gamma = 409 \pm 23 \;h \; M_{\odot}/L_{\odot}$ (which is equivalent to $\Omega_{\rm m} = 0.26 \pm 0.02$). The authors state the lensing signal of the entire cluster can be replicated by the sum of the contributions from individual galactic halos; that is, there is no additional cluster dark matter beyond that centred on galaxies, as we have assumed in our model. It is interesting to note that this result in combination with ours would imply the matter fraction not virialised into halos is $\rho_{\rm void}/ \rho  \sim 0.5$.
\par 
In summary we find our results for power law slope consistent with the literature, and the mass-to-light ratio consistent with the (albeit large) range of values quoted. 

\section{Summary and discussion}
\label{sec:summary}
In this paper, we have developed an estimator (Equation \ref{eq:normmag}) for the weak lensing convergence based on the astrometric properties of foreground galaxies. The key assumptions underlying the estimator are 
\begin{itemize}
    \item a matter density comprised of universal halo profiles superimposed on a homogeneous background,
    \item the magnification is weak,
    \item the lines of sight to SN Ia are equivalent to a random sample,
    \item the masses of dark matter halos may be estimated from galactic magnitudes using an average mass-to-light ratio.
\end{itemize}
We have demonstrated the effectiveness of the estimator by correlating it with SN Ia residuals to a best-fit Hubble diagram. Using a $\beta = 1.8$ profile with the M08 concentration model, this is $\rho = 0.166 \pm 0.046 \mbox{(stat)} \pm 0.011 \mbox{(sys)}$ for SN Ia with $0.2 < z < 1.0$. This is a detection significance of $>3\sigma$, which improves on previous results of $\sim 1.4 \sigma$ \citep{Jonsson2010, Smith2014, Macaulay2020}. Our results are not greatly affected if the NFW profile is used, or the choice of concentration model or analysis parameters within reasonable bounds. 
\par
It is natural to ask why we find a greater significance of our detection than the previous literature. Firstly, we have a $ 4 \times$ larger sample of SN Ia than \citet{Jonsson2010}, and our use of a smooth halo profile incorporating the well-established NFW profile may capture the true density profile better than a truncated isothermal sphere as used by those authors. Secondly, as we explained in Section \ref{sec:WLbyhalos}, magnification is highly sensitive to chance encounters with a low impact parameter. Although \citet{Smith2014} have similar numbers of SN Ia to us, the lensing is estimated from numbers counts of foreground galaxies only; in our method, it would be equivalent to setting $\beta = 0$ and each halo to the same mass. As we saw in Section \ref{sec:results}, lower $\beta$ has a lower correlation detection significance. While \citet{Macaulay2020} used just 196 SN Ia, better photometry from the DES platform assisted their analysis. Nevertheless, it may be challenging to use skew as a detection method. 
\par
Using Bayesian analysis, we find the mean $\beta = 1.8 \pm 0.3$ and $\Gamma = 197 ^{+64}_{-80} \, h \, M_{\odot}/L_{r, \odot}$. The SIS halo profile is ruled out at $> 95\%$ confidence. Comparing the Bayesian evidence of the power-law and NFW profiles, we find no significant difference between the two. 
\par 
In our model, $\Gamma$ and $\beta$ are uniform parameters over our entire galaxy sample, which extends between $-17 < M_r < -24$. As there may be variation by colour, luminosity, morphology and environment, they should be interpreted as a weighted population average, with the peak of weighting at $M_r \sim -21$.
\par 
We have considered the effects of Malmquist bias on our results. We show that due to magnitude limits in the SN Ia data, $\Gamma$ is likely biased low by $\sim 10\%$. This is well-within the confidence interval of our posterior, and so we do not adjust our results for this. We have also shown that the lensing dispersion is biased low in the longer redshift buckets $z>0.8$ due to magnitude limits and the small footprint of the HST surveys used in Pantheon.
\par
We show the lensing dispersion is fitted by 
\begin{equation}
\label{eq:sigmalens2}
    \sigma_{\rm lens} = (0.06 \pm 0.017) (d_{\rm C} (z)/d_{\rm C}(z=1))^{3/2} \;,
\end{equation}
which is consistent with the often cited $\sigma_{\rm lens} = (0.055 \pm 0.04) z$ \citep{Jonsson2010} for $z<0.8$ but starts to diverge higher for $z>1.0$. We can compare this number to intrinsic scatter $\sigma_{\rm int}$, which is the variation in SN Ia absolute magnitude once the light curves are standardised for colour, stretch, bias and enviroment (see Equation \ref{eq:tripp}). \citet{Scolnic2018} report that $\sigma_{\rm int} = 0.09$ for Pantheon. However, the scatter is largest for the older Low-z survey, and lowest for the newest Pan-Starrs data. Early results from the DES and Foundation SN Ia surveys, indicate the true $\sigma_{\rm int} \sim 0.07$ \citep{Brout2019}. Equation (\ref{eq:sigmalens2}) therefore shows that lensing will match intrinsic scatter at $z \sim 1.2$, which is closer than previously estimated \citep{Jonsson2010}. As a result, cosmological parameter precision will be degraded in high-z surveys (\citet{Holz2005} estimate by a factor of 3 at $z \sim 1.5$). 
\par
Based on the above, we propose that inference of cosmological parameters may be improved by in high-z surveys by modifying the Tripp estimator to include a specific term to de-lens the magnitudes as follows : 
\begin{equation}
\label{eq:trippnew}
\mu = m_B - M_B + \alpha x_1 - \beta c  + \Delta_{\rm M} + \Delta_{\rm B} - \gamma \Delta m_{\rm lens}\; .
\end{equation}
The estimator $\Delta m_{\rm lens}$ (which may be seen as an environmental variable analogous and comparable in size to the host mass correction $\Delta_M$) is given by Equation (\ref{eq:normmag}). Mean parameters given in Table \ref{tab:posteriors} can be used with $\gamma = 1$\footnote{A similar modification was proposed by \citet{Smith2014} with an estimator based on spectroscopic-only galaxies. The authors find $\gamma \sim 4$, which is driven mainly by the ratio of the number density of spectroscopic galaxies to photometric ones. However, as the spectroscopic galaxy coverage may not be uniform across a given survey, this estimator seems less practical.}. Testing this, we find for the Pantheon sample that Equation (\ref{eq:trippnew}) improves the standard deviation of the Hubble diagram residuals by $0.005$ mag for $z>0.4$.
\par
In addition to correcting magnitudes, SN Ia lensing may be used as a source of cosmological information in itself, and we will explore this in future work.

\subsection{Future surveys}
In this work, we used the Pantheon SN Ia compilation because of its uniform calibration, well-characterised bias corrections, and large overlap with the SDSS galaxy survey. However the drawbacks are the lack of a single SN Ia detection efficiency function (due to Pantheon merging several surveys) and shallow SDSS photometry. This meant the Malmquist bias was not straightforward to estimate.
\par
The Dark Energy Survey \citep{Abbott2016} offers several advantages. The associated supernova survey is expected to catalog a few thousand \textit{photometrically} classified SN Ia, down to a deep field depth of $r < 25.5$ \citep{Smith2020}. As a result of photometric classification, the detection efficiency will be simpler to model and there is less risk of a biased selection of lines of sight. In addition, the foreground galaxy catalog will be to equivalent depth as the supernova survey, and as a result bias should be small and straightforward to calculate.  DES data has been used to produce a map of lensing convergence derived from a weak lensing shear analysis \citep{Jeffrey2021}, and it will be particularily interesting to compare such maps derived from shear to those derived from magnification. 
\par 
The Rubin LSST Observatory, expected to commence survey operations in October 2023, will reach approximately 2 magnitudes deeper. It is expected catalog $\sim 10,000$ SN Ia each year \citep{Zhan2018} between $0.2 < z < 0.8$ in the southern sky. With these enlarged data sets, we anticipate improved luminosity, morphological and colour characterisation of $\Gamma(M, p)$. 
\par
Looking to space missions, the Roman Space Telescope will conduct a SN Ia search as part of its galaxy survey mission. With an optimized survey strategy \citep{Hounsell2018}, it may discover $>10,000$ SN Ia out to $z \sim 2.5$ over the course of its mission. The fractional distance modulus uncertainty per $0.1z$ bin is expected to be $4 \times 10^{-3}$ (a factor of $10\times$ improvement on the Pantheon data set). With this high-precision data set, we would expect to detect lensing at $> 15\sigma$ confidence. It will be particularily interesting to test the modified Tripp estimator we proposed in Equation (\ref{eq:trippnew}) on this data set.

\section*{Acknowledgements}

We thank Edward MacAulay, David Bacon and Sunayana Bhargava for helpful discussions on this work. We also thank the referee for helpful comments that have improved the paper. 
\par 
The Millennium Simulation databases used in this paper and the web application providing online access to them were constructed as part of the activities of the German Astrophysical Virtual Observatory (GAVO). This work has also made use of CosmoHub. CosmoHub has been developed by the Port d'Informació Científica (PIC), maintained through a collaboration of the Institut de Física d'Altes Energies (IFAE) and the Centro de Investigaciones Energéticas, Medioambientales y Tecnológicas (CIEMAT) and the Institute of Space Sciences (CSIC \& IEEC), and was partially funded by the \say{Plan Estatal de Investigación Científica y Técnica y de Innovación} program of the Spanish government.
\par 
OL and PL acknowledge support from an STFC Consolidated Grant ST/R000476/1, and PL acknowledges STFC Consolidated Grant ST/T000473/1.

\section*{Data Availability}
Upon publication, code, figure scripts and sample SQL queries will be made available at \url{https://github.com/paulshah/SNLensing}.
 



\bibliographystyle{mnras}
\bibliography{SnMagCitations} 



\appendix

\section{Specific halo profiles}
In this appendix, we summarize analytical results for the lensing magnification by certain halo profiles. 
\label{sec:haloprofile}
\subsubsection{Singular Isothermal Sphere}
The singular isothermal sphere (SIS) is derived from the assumption that dark matter haloes are thermalised with a homogeneous temperature. The density profile is
\begin{equation}
    \rho_{\rm SIS}(r) = \frac{\sigma_{v}^{2}}{2\pi G r^2} \; ,
\end{equation}
where $\sigma_{\rm v}$ is the isotropic velocity dispersion of dark matter. The mass inside $r_{200}$ is $M_{200} = \frac{800\pi}{3} \rho_{\rm c} r_{200}^{3}$ and if this is assumed to be fully virialised, we can obtain the velocity dispersion as a function of $M_{200}$ from the virial theorem as 
\begin{equation}
\label{eq:veldispSIS}
    \sigma_{\rm v}^{6} = \frac{\pi}{6} 200 \rho_{\rm c} M_{200}^2 G^3 \; .
\end{equation}
It is then straightforward to integrate $\rho_{\rm SIS}$ for the surface density, and we obtain convergence and shear as
\begin{equation}
    \kappa_{\rm SIS} = \gamma_{\rm SIS} = \frac{G \sigma_{\rm v}^2}{2\Sigma_{\rm c}}\frac{1}{b} \; ,
\end{equation}
(see for example \citet{Bartelmann2001} equation (3.19)). Hence the magnification is
\begin{equation}
\label{eq:SISmag}
\mu = \frac{\theta}{\theta_{\rm E}-\theta} \;,
\end{equation}
where
\begin{equation}
\theta_{\rm E} = \frac{4\pi \sigma_{v}^{2}}{c^2} \frac{D_{\rm ds}}{D_{\rm s}}
\end{equation}
is the Einstein radius of the halo. 
\par
We see that at large radii, $\Delta \mu = \mathcal{O}(\theta^{-1})$ and so in the case of a uniform surface density of lensing galaxies with overlapping halos, the total lensing amount does not converge as we extend our field radius. Nevertheless, we use the SIS profile as a useful control profile to compare with others that are better motivated. 

\subsubsection{$\beta = 1$}
A form of softened isothermal sphere, the convergence has a closed form expression for $\beta = 1$ which is
\begin{equation}
  \label{eq:betasurfdens1}
  \kappa_{\beta =1} = 
  \begin{cases}
     \frac{ 4 \delta_{\rm c} \rho_{\rm c} r_{\rm s}}{\Sigma_c \sqrt{(1-x^2)^2}} \mbox{arctanh}(\sqrt{\frac{1-x}{1+x}}) & x<1 \\[1em]
     \frac{ 4 \delta_{\rm c} \rho_{\rm c} r_{\rm s}}{\Sigma_c \sqrt{(x^2-1)^2}} \mbox{arctan}(\sqrt{\frac{x-1}{1+x}}) & x>1 \;. 
  \end{cases}
\end{equation}
where $x = b / r_{\rm s}$ is the dimensionless impact parameter in units of the scale radius.

\subsubsection{NFW halo}
Navarro, Frenk and White \citet{Navarro1996} (NFW) proposed a profile to empirically fit their N-body simulations of collapsed dark matter halos over a wide range of masses. The density profile is 
\begin{equation}
    \rho_{\rm NFW}(r) =  \frac{\delta_{\rm c} \rho_{\rm c}}{(\frac{r}{r_{\rm s}})(1+\frac{r}{r_{\rm s}})^2} \; .
\end{equation}
The scale radius $r_{\rm s} = r_{200} / c$ where the concentration paramater $c$ is thought to be weakly dependent on the halo mass, with smaller halos being more concentrated \citet{Navarro1996}. 
\par 
The profile is softer than the isothermal sphere at small radii, and turns over at $r_{\rm s}$ to $r^{-3}$. Although its total mass diverges logarithmically, we may equate $M_{200} = M(r_{200})$ and obtain
\begin{equation}
    \delta_{\rm c} = \frac{200}{3} \frac{c^3}{\ln{(1+c)} - c/(1+c)} \; . 
\end{equation}
\citet{Wright2000} find that
\begin{align}
\label{eq:NFWconv}
    \kappa_{\rm NFW} = 
    \begin{cases} 
    \frac{2 r_{\rm s} \delta_{\rm c} \rho_{\rm c}}{\Sigma_{\rm c} (x^2 -1)} \{ 1- \frac{2}{\sqrt{1-x^2}} \textrm{arctanh}{\sqrt{\frac{1-x}{1+x}}} \} & x<1 \\[1em]
    \frac{2 r_{\rm s} \delta_{\rm c} \rho_{\rm c}}{3\Sigma_{\rm c}}  & x=1 \\[1em]
    \frac{2 r_{\rm s} \delta_{\rm c} \rho_{\rm c}}{\Sigma_{\rm c} (x^2 -1)} \{ 1- \frac{2}{\sqrt{x^2-1}} \arctan{\sqrt{\frac{x-1}{x+1}}} \} & x>1 .
    \end{cases}
\end{align}
The shear is
\begin{equation}
\label{eq:NFWshear}
    \gamma_{\rm NFW} = 
    \begin{cases}
      \frac{ r_{\rm s} \delta_{\rm c} \rho_{\rm c}}{\Sigma_{\rm c}} g_{<}(x) & x<1,\\[1em]
      \frac{ r_{\rm s} \delta_{\rm c} \rho_{\rm c}}{\Sigma_{\rm c}} \left[ \frac{10}{3} + 4 \ln{\frac{1}{2}} \right] & x=1, \\[1em]
    \frac{ r_{\rm s} \delta_{\rm c} \rho_{\rm c}}{\Sigma_{\rm c}} g_{>}(x)& x>1 \; ,
    \end{cases}
\end{equation}
where
\begin{equation}
    \begin{split}
    g_{<} = \; & \frac{ 8\textrm{arctanh}\sqrt{(1-x)/(1+x) }}{x^2\sqrt{1-x^2}} + \frac{4}{x^2}\ln{\frac{x}{2}}  \\
    & \qquad - \frac{2}{(x^2-1)} + \frac{4\textrm{arctanh}\sqrt{(1-x)/(1+x)}}{(x^2-1)(1-x^2)^{1/2}} , \\
    g_{>} = \; & \frac{8\textrm{arctan}\sqrt{(x-1)/(1+x) }}{x^2\sqrt{x^2-1}} + \frac{4}{x^2}\ln{\frac{x}{2}} \\
    & \qquad - \frac{2}{(x^2-1)} + \frac{4\textrm{arctan}\sqrt{(x-1)/(1+x)}}{(1-x^2)^{3/2}}  . \\
    \end{split}
\end{equation}

\subsubsection{Hernquist profile}
For the Hernquist profile ($\beta = 3$) the convergence is
\begin{align}
\label{eq:betasurfdens3}
    \kappa_{\rm Hern} = & \frac{\delta_{\rm c} \rho_{\rm c} r_{\rm s}}{\Sigma_c (1-x^2)^2} ((2+x^2)S(x) -3) \;\;, \\[1em]
    S(x) = &
    \begin{cases} 
        \frac{1}{1-x^2} \log{(1+ \sqrt{(1-x^2)/x})} & x<1 \;\;,\\[1em]
        \frac{1}{x^2-1} \arccos{1/x} & x>1 \;\;,
    \end{cases}
\end{align}
as given by \citet{LarsHernquist1990}.

\section{Analytic derivation of $\sigma_{\rm lens}$}

Gravitational lensing is a 2-body interaction. We may estimate the interaction rate per source as $\propto n_c V$ where where $n_c$ is the comoving number density and $V$ is an applicable comoving volume. Therefore, we may expect $\sigma_{\rm lens} \propto V^{1/2}$ if $n_c$ is approximately constant and galaxies are randomly distributed. The applicable volume should be related to the lensing efficiency squared (that is, the \say{cross-section} of the interaction) integrated over the distance to the source. 
\par
A general formula for the variance of the lensing convergence due to halos has been derived in \citet{Kainulainen2009, Kainulainen2011}, who used the same density model as we do. We summarize their derivation, and explicitly integrate it for the case of NFW haloes. 
\par
It is convenient for this calculation to split the convergence as 
\begin{equation}
    \kappa \equiv \kappa_H + \kappa_E \;\;,
\end{equation}
where
\begin{equation}
    \kappa_E = -\int_{0}^{r_s} \frac{3 \Omega_{M,0}}{2c^2} H_0^2 (1+z(r)) d_{\rm lens}(r) dr \;\;
\end{equation}
is the \say{empty beam} value corresponding to maximum de-magnification, which is constant for a given source distance. $d_{\rm lens} \equiv \frac{r (r_s - r)}{r_s}$ is the lensing efficiency for a source at comoving distance $r_s$ and a lens at $r$. The convergence due to matter halos along a given line of sight to a source may be written as the sum of contributions discretised over $N_S$ bins in comoving distance $\{r_i\}$ and $N_R$ bins in comoving impact parameter $\{b_m\}$ :  
\begin{equation}
\label{eq:kappah}
\kappa_H = \sum_{i=1}^{N_S} \sum_{m=1}^{N_R} k_{im} \kappa_{1,im} \;\;.
\end{equation}
The $k_{im}$ are random number counts of lensing halos in the comoving volume defined by the interval $(r_i, r_i+\Delta r_i)$ and $(b_m, b_m +\Delta b_m)$ where the binning is arbitrary, but small enough such that the convergence $\kappa_{1,im}$ of single halos can be taken to be a fixed value over the bin. 
\par
The $k_{im}$ have Poisson statistics
\begin{align}
    P(k_{im}) & \sim  \mbox{Poisson}(\Delta N_{im}) \\
    & =  \frac{(\Delta N_{im})^{k_{im}}}{k_{im}!} \exp{-\Delta N_{im}} \;\;,
\end{align}
where the Poisson parameter is the expectation of the number of halos in each bin 
\begin{equation}
    \Delta N_{im} = n_c (2\pi b_m \Delta b_m) \Delta r_i \;\;.
\end{equation}
Photon conservation is ensured by the requirement that the matter density of halos averages to the homogeneous matter density $\Omega_{\rm m}$.

\subsection{$\sigma_{\rm lens}$ for a general halo profile}
For simplicity here, we take all halos have the same mass and parameters - an extra bucketing scheme can be easily introduced to generalise this if desired. We also assume that comoving number density of halos is constant with time, the matter distribution along lines of sight to supernovae has the same distribution as randomly drawn lines of sight, and the Poisson numbers $k_{im}$ are uncorrelated.
\par
The assumption of a randomly drawn and unbiased line of sight, while consistent with the \say{stochastic} treatments of \citet{Holz2005, Jonsson2010, Kainulainen2009}, is not trivial : SN Ia are not located randomly in empty space but instead in galaxies. To the extent that galaxies cluster (that is, exhibit positive spatial correlation of their number density), SN Ia may be expected to lie preferentially in over-dense regions, and the Poisson numbers $k_{im}$ will indeed be correlated. \citet{Kainulainen2011a} examined spatial correlations in this model by using the halo model, which splits the contribution into 1-halo (peak) and 2-halo (background) components, and found the additional contribution to the variance due to the 2-halo term was relatively small. However, taking into account mass variability and potential halo substructure, there formulae stated here should be seen as a \textit{lower bound} for the true lensing dispersion.  
\par 
Equation (\ref{eq:kappah}) is the weighted sum of uncorrelated (but not identically distributed) Poisson random numbers $k_{im}$. We can therefore write 
\begin{equation}
    \mbox{Var} (\kappa) = \sum_{im} \kappa_{1,im}^2 (z_s) \Delta N_{im}
\end{equation}
where we have used the properties of variance that Var$(X_i + c)$ = Var$(X_i)$ for $X_i$ any random variable and $c$ constant, Var$(a_i \sum X_i)$ = $\sum a_i^2 \mbox{Var}(X_i)$ for any uncorrelated random variables, and specifically for the Poisson distribution Var$(k_{im}) = \Delta N_{im}$. 
\par 
Converting the sum to an integral we have (see also equation (70) of \citet{Kainulainen2011}) 
\begin{equation}
\label{eq:siglens}
\begin{split}
    \sigma_{\rm lens}^2 = 2\pi n_c  & \left[ \frac{3}{2} \Omega_{m,0} \frac{H_0^2}{c^2} \right]^2 \int_{0}^{r_s} dr \;d_{\rm lens}^2(r,r_s) (1+z(r))^2\\
    & \int_{b_{\rm min}}^{b_{\rm max}} b db \left( \int_{b}^{b_{\rm max}} \frac{2x \;dx}{\sqrt{x^2 - b^2}}  \frac{\rho(x,r)}{\bar{\rho}_m} \right)^2 \;\;,
\end{split}
\end{equation}
where $b_{\rm min}$ and $b_{\rm max}$ are arbitrary comoving cutoff radii imposed to regularize the integrals ($b_{\rm max}$ may be taken to be a truncation radius). The latter integral is the comoving halo surface density at impact parameter $b$ normalised to units of the average matter density $\bar{\rho}_m$. We see that the volume element enters via the product $d_{\rm lens}^2 dr$.

\subsection{$\sigma_{\rm lens}$ for the NFW halo profile}
We use the formula for $\kappa_{\rm NFW}$ specified in Appendix \ref{sec:haloprofile} and define $f_c = 1/ (\log{(1+c)} - c/(c+1)) $. In the limit $x \gg 1$ we  have 
\begin{equation}
    \kappa_{1, im} \simeq \frac{2G M_{200} f_c}{c^2} d_{\rm lens}(r_i, r_s) \frac{1}{b_m^2} \;,
\end{equation}
and in fact this will be an adequate proxy for our purposes. $c$ in the denominator is the speed of light. The proportionality to $1/b^2$ is due to $\rho(r) \propto 1/r^3$ for the NFW halo at large $r$. 
\par
\begin{figure}
    \centering
    \includegraphics[width=\columnwidth]{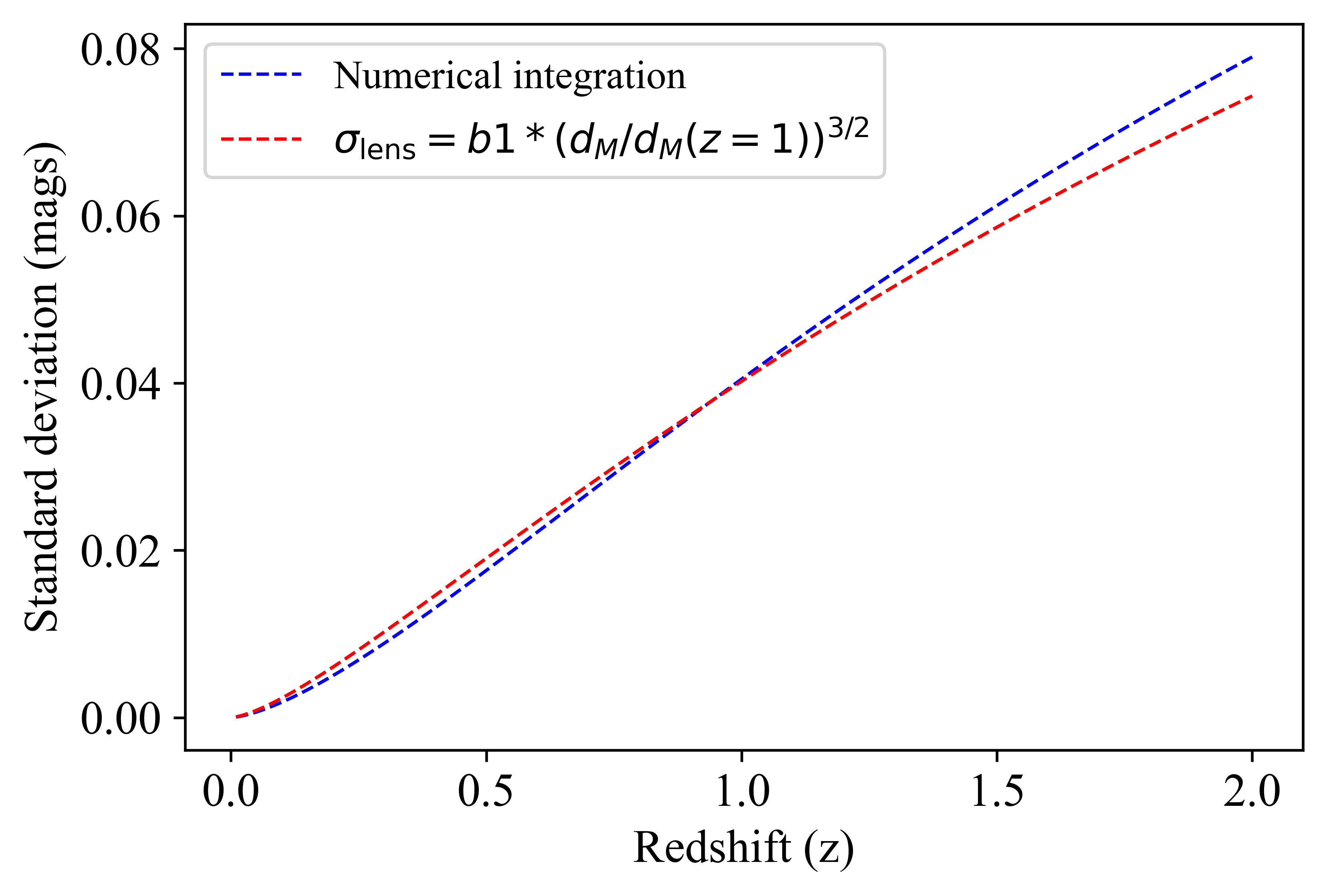}
    \caption{$\sigma_{\rm lens}$ computed by numerical integration with parameters $M_{200}=10^{11} M_{\odot}$, $c = 6$, $n_c = 2.5$ Mpc$^{-3}$, together with the approximate formula \ref{eq:sla} for $b_{\rm min} = r_{sc}$.}
    \label{fig:sigmalens}
\end{figure}

Substituting this into equation (\ref{eq:siglens})
\begin{equation}
    \sigma_{\rm lens}^2 = A \times I
\end{equation}
where the constant of proportionality 
\begin{equation}
    A = (\frac{2G M_{200} f_c}{c^2})^{2} 2\pi n_c \;\;,
\end{equation}
and 
\begin{equation}
\label{eq:sl}
 I = \int_{0}^{r_s} dr \int_{b_{\rm min}}^{b_{\rm max}} db \frac{r^2 (r_s - r)^2}{r_s^2} \frac{1}{b^3} \;\;.
\end{equation}
All variables are expressed as comoving distances. With the final assumption that $b_{\rm max} \gg b_{\rm min}$ are fixed and not functions of $r$, we arrive at
\begin{equation}
\label{eq:sla}
    \sigma_{\rm lens} \simeq  \sqrt{\frac{A}{60}} \frac{r_s^{3/2}}{b_{\rm min}} \;\;.
\end{equation}
In fact the integral without assuming the large $x$ approximation can be done numerically, in which case we find $b_{\rm min} \simeq r_{sc} = r_{200}/c$, as shown in Figure \ref{fig:sigmalens}.

The behaviour $\sigma_{\rm lens} \propto r^{3/2}$ is generic provided our assumptions hold, and we checked our formula matches that of a randomly-generated galaxy catalog. We have also checked the formula against $\sigma_{\rm lens}$ calculated from a galaxy catalog generated from the Millenium simulation by \citet{Henriques2012}. There is a modest extra variance, increasing with distance, indicating the spatial correlation of galaxies. However, when centering the lines of sight on random \textit{galaxies}, we find $\sigma_{\rm lens}$ is larger than our formula by $\sim 50\%$ at $z =1$. 
\par 
It is this last point that has interesting implications for supernova cosmology : the Pantheon SN Ia may be special if they lie preferentially where matter density and clustering is different to the background average.


\bsp	
\label{lastpage}
\end{document}